\documentclass[twocolumn,showpacs,preprintnumbers,amsmath,amssymb]{revtex4}

\usepackage{graphicx}
\usepackage{dcolumn}
\usepackage{bm}

\begin{document}

\preprint{}

\title{Peristaltic modes of single vortex in 
the Abelian Higgs model}

\author{Toru Kojo, Hideo Suganuma and Kyosuke Tsumura}

\affiliation{
Department of Physics, Kyoto University, Kyoto 606-8502, Japan 
}

\date{\today}

\begin{abstract}
Using the Abelian Higgs model,
we study the radial excitations of single vortex
and their propagation modes along the vortex line.
We call such beyond-stringy modes 
{\it peristaltic} modes of single vortex.
With the profile of the static vortex, 
we derive the vortex-induced potential, i.e., 
single-particle potential for the Higgs 
and the photon field fluctuations around the static vortex, 
and investigate the coherently propagating fluctuations
which corresponds to the vibration of the vortex. 
We derive, analyze and numerically solve the field equations of
the Higgs and the photon field fluctuations around
the static vortex with various Ginzburg-Landau parameter $\kappa$
and topological charge $n$.
Around the BPS value or critical coupling $\kappa^2$=1/2,
there appears a significant correlation between
the Higgs and the photon field fluctuations mediated by the static 
vortex.
As a result, for $\kappa^2$=1/2,
we find the characteristic new-type discrete pole
of the peristaltic mode corresponding to the
quasi-bound-state of coherently fluctuating fields and the static vortex.
We investigate its excitation energy, correlation energy of 
coherent fluctuations,
spatial distributions, and the resulting magnetic flux behavior in detail.
Our investigation covers not only usual Type-II vortices with $n$=1
but also Type-I and Type-II vortices with $n\in Z$
for the application to various general systems 
where the vortex-like objects behave as the 
essential degrees of freedom.
\end{abstract}


\maketitle
\section{\label{sec:intro}Introduction}

There have been wide interests in the study of 
the topological objects such as the vortices \cite{abrikosov}, 
monopoles \cite{monopole, monopole2},
and instantons \cite{polyakov} because of their importance for understanding 
some crucial aspects of the nonlinear field theories \cite{Raja, man}.
They are not included in the original action manifestly,
but appear as the localized finite energy solutions of the 
nonlinear field equations.
These solutions can be classified with the topological number 
which comes from the topological property of the system.
Since the topological objects have spatially localized,
finite energy  configurations and some topologically 
conserved charge,
it is fascinating to regard them as dynamical degrees 
of freedom in the system.

One of such objects is the Abrikosov vortex \cite{abrikosov} 
in Ginzburg-Landau (GL) theory,
which appears as the magnetic flux squeezed
by the Cooper-pair condensate. 
When the external magnetic field applied to the superconductor 
exceeds some critical strength, 
the Cooper-pair is dissociated and 
the node of the Cooper-pair forms a vortex line.
The single-valued property of the Cooper-pair 
around vortex lines leads
the quantization of the total magnetic flux,
as $e\Phi=2\pi n\ (n \in {\bf Z})$, where
$e,\Phi, n$ represent the effective gauge coupling of the Cooper-pair,
the total magnetic flux and the topological number, respectively.
The topological number $n$ characterizes the
topology of the system
and we can classify vortices in terms of $n$.  

There exist many applications employing the concept of the Abrikosov vortex 
and its relativistic version, the Nielsen-Olesen vortex \cite{Nielsen}.
One example is the squeezed color-electric flux between the 
quark and anti-quark in QCD.
Nambu \cite{Nambu}, 't Hooft \cite{'t Hooft} 
and Mandelstam \cite{Mandel} discussed that
there exist the vortices which appear as the color-electric flux tubes
squeezed by the condensation of magnetic monopoles.
As a result of the color-electric flux squeezing,
we can only observe colorless mesons and baryons, 
which are color singlet bound states of the colored quarks
with the bond of the color-electric flux.
This scenario is a possible explanation 
for ``color confinement'', which is
the experimental fact in hadron physics
\cite{confinement}.
The lattice QCD Monte Carlo calculations \cite{rothe} for
${\rm Q\bar{Q}}$ \cite{creutz} and 3${\rm Q}$ \cite{takahashi,ichie} 
potentials strongly support the flux tube picture,
and show the universality of the string tension of the flux-tube, 
$\sigma\simeq$ 0.89 GeV, in both ${\rm Q\bar{Q}}$ and 3${\rm Q}$ cases.

An another example is the cosmic string in astrophysics
as a seed of the galaxy formation and baryogensis \cite{vile}.
Kibble \cite{kibble} and Zurek \cite{zurek} discussed 
the distribution of the topological defects and their cosmological 
evolution after the quench of the early Universe and the resulting 
phase transition. 
They argued that the cosmic strings as topological defects
would produce a characteristic signature in the cosmic microwave
background.
Following their scenarios, the phase transition dynamics are
closely investigated in the numerical simulations
using the phenomenological 
time-dependent GL (TDGL) theory \cite{tink} in both cases with 
global symmetry and local gauge symmetry.
In particular, in the case of the gauge symmetry,
it is discussed that
the initial thermal magnetic fluctuations just after the quench 
play important roles on the later distribution of the vortices 
such as the clusters of vortices
\cite{hind,step}.
These investigations are closely correlated with the 
condensed matter physics in a viewpoint to understand
the phase transition dynamics,
and provide the interesting subjects to link 
different fields of research.

Now we turn to the dynamical aspects of single vortex as
a ``pseudo-particle''. 
For the treatment of the vortex motion,
the vortex is usually regarded as one dimensional (1-D) thin object 
like a string,
neglecting the extension perpendicular to the vortex line. 
Such a treatment is expected to be sufficient when
the length of the string is much larger than the length scale
in the radial direction.
This condition comes from the fact that
the excitation energy of string as a 1-D object is 
typically $\sim \pi/l_s$ ($l_s$ is the string length).
Then, for the vortex with large length,
the stringy excitation is more important than 
radial one with changing its thickness.
Even if the above condition is not well-satisfied,
we have only to consider extensive modes of the magnetic flux
when the Type-II nature is very strong, i.e.,
the Ginzburg-Landau parameter $\kappa=\delta / \xi$ is very large
(Here $\delta$ is the penetration depth of the magnetic field
and $\xi$ is the coherence length of the Higgs field).
This is because when Type-II nature is very strong,
the topological defect of the Higgs field along the 
vortex line is strongly squeezed and its extension modes 
perpendicular to the vortex line cost a large energy.
Then, in the Type-II limit of the vortex, 
the main excitation is given by the stringy modes, and 
the subject is linking to one of the most fundamental problems 
in the quantum string theory\cite{P98}, 
which gives not only a candidate of the grand-unification 
including gravity but also a new method to analyze nonperturbative QCD \cite{NSK07}.
So far, there are so many interesting studies done on the quantum string.
For instance, the quantum fluctuations of the string lead to 
a ``roughening", i.e., a widening of the string \cite{LMW81}, and 
several studies of quantum Nielsen-Olesen strings \cite{O94,ACPZ96} 
showed that quantum vibrational modes make the 
vortex world-sheets crumpled and dominated by ``branched-polymers" 
as a disease of quantum strings\cite{O94}. 

However, we sometimes encounter the cases 
where the vortex cannot be regarded as a stringy object.
%
We are interested in such cases.
One example is the Type-II superconductors 
in condensed matter physics near 
$\kappa^2$=1/2, i.e., Bogomol'nyi-Prasad-Sommerfeld (BPS) \cite{bogo,PS} 
value or critical coupling.
In such cases, the squeezing of the magnetic flux due to Cooper-pair
condensation is not very strong
and both the Cooper-pair and photon fluctuations 
can play important roles.
Moreover, their interplay
may provide interesting phenomena.

An another interesting example is the quark systems
such as the meson and baryon which
have the color-electric flux tube with the total length
about 1 fm,
which is not very large in comparison with the flux tube extension of
about 0.4 fm.
In addition, according to the analysis
using the dual-Ginzburg-Landau model \cite{Suganuma, Suganuma2},
Type-II nature also seems to be not so strong,
and then the radial motion of the vortex can play 
an important role.
Studies on such excitation modes are important 
to understand the structure of hadrons.
For the gluonic excitations, the lattice QCD studies for 
heavy ${\rm Q\bar{Q}}$ \cite{juge} and 3${\rm Q}$ \cite{taka2} system 
both provide the 1st gluonic excitation energy as
${\rm \triangle}E \sim$ 1 GeV in the
the total length of the flux, $L_{\rm min}$=0.5-1.0 fm, as the typical
hadronic scale.
Small dependence on $L_{\rm min}$ 
provides the possibility to understand the 1st gluonic 
excitation as not simple stringy vibration
but the radial excitation with the fixed boundary
effects on the flux-tube,
because the simple stringy mode is likely
to behave proportionally to 1/$L_{\rm min}$. 
The studies on the energy of the radial excitation  
with comparing it to that of lattice results
will give a qualitative explanation for the gluonic excitations
in the case of relatively small but not too short $L_{\rm min}$.
There seem to exist several other cases where
the radial excitations play important roles.

\begin{figure}[b]
\vspace{-0.3cm}
 \includegraphics[width=8.5cm]{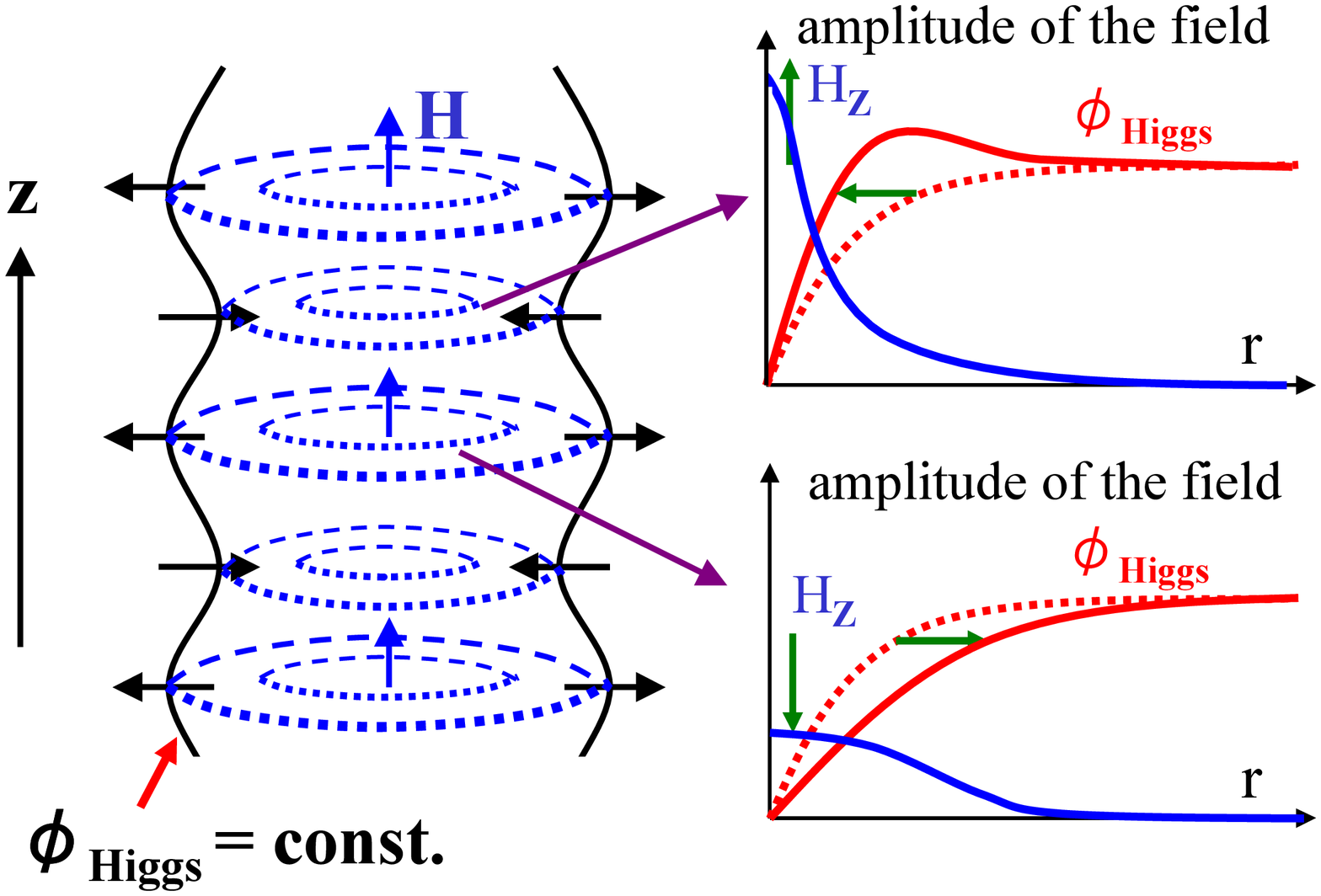}
 \vspace{-0.0cm}
      \caption{\label{fig:peristic}
\small{ The illustration for the {\it peristaltic} mode of the vortex. 
The axial-symmetric oscillation
perpendicular to the vortex line propagates
with the excitation of the Higgs and photon fields.
The waving solid line in the left side represents the surface
where the amplitude of the Higgs field 
equals some constant value.
The right hand side represents the amplitude of the
Higgs and the magnetic fields in the cross section 
perpendicular to the vortex line.
Through this process, the conserved total magnetic flux 
repeatedly contracts and relaxes.} }
\end{figure}

Putting applications to the above examples in perspective,
we will discuss the axial-symmetric 
radial excitation modes 
and their propagation along the vortex line.
Hereafter, we will call this excitation mode {\it peristaltic} mode.
The schematic picture of the peristaltic mode is shown in
Fig.\ref{fig:peristic}.
To discuss the peristaltic modes, we utilize the 
Abelian Higgs action which includes
the Higgs and the photon fields.
In terms of this action, the peristaltic modes
are characterized by the Higgs and photon field fluctuations.
In particular, the time-dependent photon fluctuations,
which are sometimes neglected, 
play important roles on the corporative behavior with 
the Higgs field fluctuations.
These fluctuations are regarded as the vibration of the vortex
from the viewpoint that the  
vortex is the dynamical degrees of freedom in the system.

In this work, since we concentrate on the radial excitation modes,
we simply neglect the effects of the boundary of the vortex edge, considering
the vortex with the infinite length.
The bending of the vortex is also neglected.
We investigate both Type-I and Type-II vortices 
with not only usual $n$=1 but also $|n| \ge 2$
for the future applications to 
the non-equilibrium or multi-vortices systems 
where the giant vortices can appear
as excited states.
In the Type-I case,
if the system contains only finite magnetic flux
in contrast to the equilibrium case usually realized in the laboratory,
these flux gather and become a $|n| \ge$ 2 giant vortex 
because Type-I vortices attract each other.
On the other hand, in the Type-II case,
the giant vortex is unstable and splits into the small vortices.
In spite of this, when $\kappa^2$ is not too large,
it is worth considering the vibration of the giant vortex
because the gradient of the vortex-vortex interaction
is small for the small separation of vortices \cite{jacob}
and we can expect that the fission process develops slowly.

The organization of this paper is as follows.
In Sec.II we briefly review the treatments of the static profile
of single vortex.
In Sec.III we derive the effective action 
and the resulting equation of motion for 
the fluctuations around the vortex.
The effects of the static vortex background on the fluctuations
are emphasized.
In Sec.IV we show the numerical results for the excitation modes 
of the $n$=1 vortex with 0.1$\le$$\kappa^2$$\le $1.0.
Based on the classification with $\kappa^2$,
we closely discuss their excitation energy,
correlation energy between the Higgs and photon fields,
spatial distribution of fluctuations,
and the resulting magnetic flux behavior. 
It will be shown that, in the case of $\kappa^2 \simeq$ 1/2, 
the interplay of the Higgs and the photon fields
plays an essential role for the existence of 
the new-type discrete pole.
We also discuss the excitation modes around  $|n| \ge $ 2 vortices.
In Sec.V we summarize this paper with
the perspective for the future direction.

\section{\label{sec:U(1))} Abelian Higgs model and the static Vortex}

In this section, we briefly review the 
treatments of the static vortex in terms of
Abelian Higgs model \cite{Nielsen}.
This type of action is very popular
and has been used to describe 
cosmic string in the astrophysics \cite{vile},
the color flux tube in QCD \cite{Suganuma2}, and so on.
We use this Lorentz invariant action
to investigate the properties of the vortex as a ``pseudo-particle''.

For the finite temperature cases,
one usually uses the phenomenological model  
with replacing the 2nd time-derivative in Abelian Higgs model 
with the 1st derivative
although the microscopic foundation of this treatment 
is still not conclusive.
The noise term is sometimes further added.
These replacements change the equation of motion
from the wave equation to the diffusive equation
which respects the dissipation effects due to the thermal fluctuations.
We will discuss the finite temperature case elsewhere.

\subsection{The action and the field equation}

Let us consider the ideal system 
which contains only one
vortex with infinite length along the $z$-axis.
Our purpose is to investigate the excitation mode
of the vortex in such a system.
We start with the following action 
with natural units ($c=\hbar=1$)
\begin{eqnarray} \label{action1}
\bar{S} = \int d^4\bar{x} \bigg\{ \mid ( \bar{\partial}^{\mu}
 - ie \bar{A}^{\mu} )
 \bar{\psi} \mid^2
- \frac{1}{4}(\bar{\partial}^{\mu} \bar{A}^{\nu} 
- \bar{\partial}^{\nu} \bar{A}^{\mu} )^2
\nonumber \\
- \lambda( \mid \bar{\psi} \mid^2 - v^2 )^2 \bigg\},
\end{eqnarray}
where $\bar{\psi}$ and $\bar{A}_{\mu}$ represent 
the Higgs field and photon
field respectively. Here $e, \lambda, v$ are
the effective gauge coupling constant, the strength of self-coupling,
and the vacuum expectation value of the Higgs field, respectively.
$\bar{x}^{\mu}$ denote space-time coordinates and 
$\bar{\partial}^{\mu}\equiv \frac{\partial}{\partial \bar{x}_\mu}$. 
Through this paper, we describe the variables in physical unit
with bar.
For later convenience, we rescale the fields and the coordinates 
to make them dimensionless quantities, 
\begin{eqnarray}\label{rescale}
\psi = \bar{\psi}/v,\ A^{\mu} =  \bar{A}^{\mu}/v,\ 
 x^{\mu} =  ev \bar{x}^{\mu}.
\end{eqnarray}
We also rewrite the action
\begin{eqnarray}
\label{reaction}
S = e^2 \bar{S}.
\end{eqnarray}
The energy relation between the physical and our rescaled unit
can be obtained with noting the following relation, 
\begin{eqnarray}
\bar{S} = \int d\bar{t}\ \bar{L} 
\equiv \frac{1}{e^2} \int dt\ L = \int d\bar{t}\ \frac{v}{e} L.
\end{eqnarray}
Here $\bar{L}$ and $L$ are Lagrangian in physical and rescaled unit,
respectively, and they are related as $\bar{L}=\frac{v}{e} L$.
Similarly, we can obtain the energy and momentum in the physical unit
with multiplying $v/e$ to our rescaled energy and momenta.

After these rewriting, the action reads
\begin{eqnarray} \label{action2}
 S = \int d^4x \bigg\{ \mid ( \partial^{\mu} - i A^{\mu} ) \psi \mid^2
   - \frac{1}{4}(\partial^{\mu} A^{\nu} - \partial^{\nu}A^{\mu} )^2
\nonumber \\
   - \kappa^2( \mid \psi \mid^2 - 1 )^2 \bigg\}.
\end{eqnarray}
Here we have introduced the GL parameter $\kappa^2 =
\lambda/e^2 = \delta/\xi$, which is used to distinguish 
the Type-I ($\kappa < 1/\sqrt{2}$\ ) 
and Type-II ($\kappa > 1/\sqrt{2}$\ ) superconductors.
The critical case of $\kappa=1/\sqrt{2}$ is called as 
the BPS case, which lies between Type-I and Type-II.
As a remarkable fact in the BPS case, 
the photon mass $m_A$ and the Higgs mass $m_\phi$ coincide as $m_A=m_\phi$, 
because of $m_A=\sqrt{2} ev$ and $m_\phi=2\sqrt{\lambda}v$, i.e.,  
$2 \kappa^2=2 \lambda/e^2=m_\phi^2/m_A^2$,
which can be derived from the original action (\ref{action1}).

In this work, we will consider the system which is translation invariant
in the $z$-direction and contains
single vortex with infinite length along the $z$-axis.
We also impose the axial-symmetry around the $z$-axis
on the static solution.
With cylindrical coordinates ($r,\theta,z$),
we denote the complex field $\psi(x)$ in the polar-decomposed form
\begin{eqnarray}
\label{polar}
\psi(x) = \phi(t,r,\theta,z) e^{in\theta + i\chi(t,r,\theta,z)}
\ \  (-\pi\le\chi<\pi),
\end{eqnarray}
where $n$ is the topological number ($n$=0,\ $\pm$1,\ \dots) 
related to the phase of the Higgs field wave function around the $z$-axis.
For single vortex with the topological number $n\in{\bf Z}$,
the sign of $n$ is not relevant because of charge conjugation
symmetry of the Abelian Higgs model, and therefore we have only to 
consider $n$=0, 1, \dots cases without loss of generality.
Since the Abelian Higgs model has the U(1) local gauge symmetry,
we have the degrees of freedom to choose the gauge.
For the treatment of the vortex solution, the familiar gauge is
the $A_0 = 0$ gauge. 
Instead, we adopt the gauge which 
eliminates the small amplitude of the phase, 
\begin{eqnarray}
\chi = 0,
\end{eqnarray}
to simplify the equations of fluctuations in the later analysis.
For $n$=0, this gauge fixing corresponds to the unitary gauge. 

From the rotational symmetry and translational invariance
along the $z$-direction,
we adopt the following Ansatz for the static solutions
\begin{eqnarray}
\label{ansatz}
\psi(x) &=& \phi(r) e^{in \theta}, \nonumber \\
A^{\mu}(x) &=& (0,\ - A_\theta (r)\ {\rm sin} \theta,\ 
   A_\theta (r)\ {\rm cos}\theta,\ 0\ ),
\end{eqnarray}
and the static electric and magnetic fields are obtained as
\begin{eqnarray}
{\bf E}(r) = {\bf 0},\ \ 
{\bf H}(r) = (0,\ 0,\ \frac{1}{r}\frac{d}{dr}( r A_\theta )\  ).
\end{eqnarray}
As seen soon later in Eq.(\ref{sta1}) and (\ref{sta2}),
our gauge fixing and Ansatz for the static fields reproduce
the same results for the static field solutions as those in
usual $A_0=0$ gauge.

The total static energy of the system (in the rescaled unit) is obtained as
$E = -S$ with putting the time derivative terms zero,
\begin{figure*}[floatfix]

 \begin{minipage}{\linewidth}
     \includegraphics[width=15.0cm]{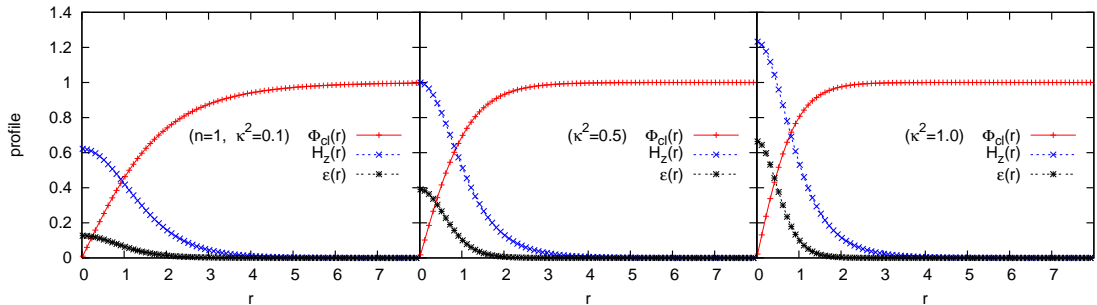}
 \vspace{-0.2cm}
 \caption{\label{profile1}
\small{ The profile for the Higgs field $\phi_{ {\rm cl} }(r)$,
  the magnetic field $H_z(r)$ and static field energy ${\cal E}(r)$ 
  for $\kappa^2$=0.1 (Type-I), $\kappa^2$=0.5 (BPS),
  $\kappa^2$=1.0 (Type-II) in the case of $n$=1 vortex.} }
\end{minipage}
\end{figure*}
\begin{eqnarray}
E_{{\rm static}} = \int\!\! dz dr
\ 2\pi r {\cal E}[\phi,A_\theta],
\end{eqnarray}
where
\begin{eqnarray}
\label{ene}
{\cal E}[\phi, A_\theta] = 
&& \!\!\! (\partial_z \phi)^2 + (\partial_r \phi)^2 
   \nonumber \\
   && \!\!\!  
   + \frac{1}{2}(\partial_z A_\theta )^2 
   + \frac{1}{2}(\partial_r A_\theta + \frac{ A_\theta }{r} )^2 \nonumber \\
   && \!\!\! + \frac{ ( n- r A_{\theta} )^2 } {r^2} \phi^2 + \kappa^2 (\phi^2 -1)^2.
\end{eqnarray}
(The full action and the energy functional are given in 
Appendix \ref{appendixa}.)
Under the above circumstances,
we consider the static vortex solutions 
$\phi(r)$ and $ A_{\theta} (r)$
which minimize the static energy.
From the variational principle
\begin{eqnarray}
\frac{\delta E_{{\rm static}} } {\delta \phi(r) }
= \frac{\delta E_{{\rm static}} } 
{\delta A_\theta (r) } =0,
\end{eqnarray}
and then the static field equations read
\begin{eqnarray}
&& \hspace{-1.0cm} -\frac{1}{r}\frac{d}{dr}\big( r\frac{d\phi}{dr} \big)
  + \frac{( n - r A_\theta )^2}{r^2}\phi 
  + 2\kappa^2 \phi (\phi^2-1) = 0, \label{sta1}\\
&& \hspace{-0.75cm} \frac{d}{dr}\big( r\frac{d A_\theta}{dr} \big)
  - \frac{A_\theta}{r} + 2( n - r A_\theta ) \phi^2 = 0 \label{sta2}.
\end{eqnarray}
Since we are interested in the vortex solution,
we consider the boundary condition (B.C.)
which is appropriate to describe the vortex at the center.
For $r\rightarrow 0$,
\begin{eqnarray} \label{bc1}
\phi (r) \rightarrow {\rm const.} \times r^n,\ 
A_\theta (r) \rightarrow {\rm const.} \times r,
\end{eqnarray}
and for $r\rightarrow \infty$,
\begin{eqnarray}\label{bc2}
&&\phi (r) \rightarrow 1 + {\rm const.} \times  r^{ -\frac{1}{2} } e^{
          -2 \kappa r},\nonumber \\
&& A_\theta (r) \rightarrow \frac{n}{r} 
 + {\rm const.} \times r^{ - \frac{1}{2}}
e^{-\sqrt{2} r}.
\end{eqnarray}
These boundary conditions are obtained
by the asymptotic analysis of the static field equations
which satisfy the following physical requirements 
to describe the vortex at the center:
(I)  at $r=0$ the vacuum expectation value
of Higgs field takes zero.
(II) the static field energy $U$ asymptotically approaches 
to the vacuum energy when $r\rightarrow
\infty$. 
Note that the first condition represents the existence of the vortex
at $r=0$
and the second one represents the magnetic fields are 
completely screened in the region far from the center of the vortex. 
The boundary conditions (\ref{bc1}) and (\ref{bc2}) 
also lead the quantization of the 
total magnetic flux $\Phi_0$,
\begin{eqnarray}
\Phi_0 \equiv \int_0^{\infty}\!\!\! dr\ 2\pi r H_z(r)
 &=& \int_0^{\infty}\!\!\! dr\ 2\pi \frac{d}{dr}( r A_\theta(r) ) \nonumber \\
 &=& 2\pi n. \label{totalflux}
\end{eqnarray}
($\Phi_0$ corresponds to $e\Phi$ in the physical unit.)

We can numerically solve the static field equations 
(\ref{sta1}) and (\ref{sta2}) with the boundary conditions
(\ref{bc1}) and (\ref{bc2}) and obtain the static solutions 
$\phi^{ {\rm cl} } (r)$ and $A^{ {\rm cl} }_\theta (r)$
for the Higgs field $\phi(r)$ and the photon field
$A_{\theta}(r)$.
Fig.\ref{profile1} shows the static profiles of the Higgs field
$\phi^{ {\rm cl} }(r)$, the magnetic field 
$H^{ {\rm cl} }(r) = \frac{1}{r}\frac{d}{dr}{(rA^{ {\rm cl} }_\theta)}$,
and the resulting static energy distribution ${\cal E}(r)$
in the vortex with various $\kappa$.

\subsection{The vortex mass and the classification of Type-I and Type-II
  superconductors}
\begin{figure}[h]
     \includegraphics[width=7.0cm]{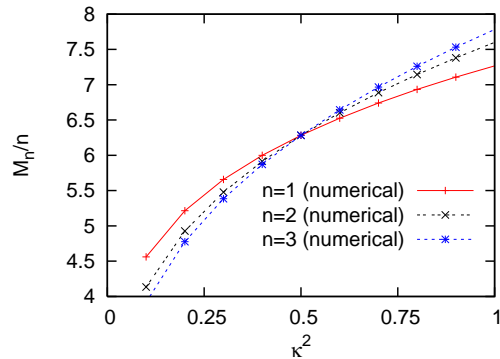}
 \vspace{-0.3cm}
      \caption{\label{fig:masssmall}
\small{ The plot for $\kappa^2$ dependence 
of the vortex mass divided by topological number $n$, $M_n/n$ 
for $n$=1, 2, 3 in the rescaled Abelian Higgs model in units of Eq.(\ref{rescale}).} }
\end{figure}

\begin{figure}[h]
     \includegraphics[width=7.0cm]{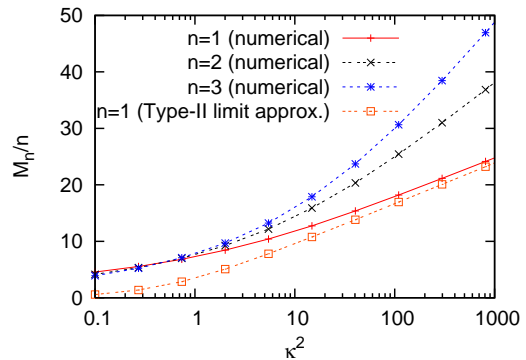}
 \vspace{-0.3cm}
      \caption{\label{fig:vortexmass}
\small{ The plot for $\kappa^2$ dependence 
of the vortex mass divided by topological number $n$, $M_n/n$ 
for $n$=1, 2, 3 in the rescaled Abelian Higgs model in units of Eq.(\ref{rescale}).
The line with circle represents the mass of the $n$=1 vortex
calculated with Type-II limit (large $\kappa^2$) approximation 
in \cite{abrikosov}.
The horizontal axis represents the log plot of $\kappa^2$ value 
in the much wider range than that in Fig.~\ref{fig:masssmall}. } }
\end{figure}

Using the distribution of the static field energy, 
we can calculate the vortex mass (with quantum number $n$) 
per unit length in the $z$-direction, i.e., 
\begin{eqnarray}
M_n \equiv 2\pi \times 
\int_0^{\infty} \!\!\! dr\ r 
{\cal E} [ \phi = \phi^{ {\rm cl} } , 
A_\theta = A^{ {\rm cl} } ].
\end{eqnarray}
Note that by definition of the action, the origin of energy is
zero when the system has no vortex.
The vortex mass is classified in the following way,
\begin{eqnarray} 
M_n &<& 2\pi n\ \ \ \ (\kappa^2 < 1/2\ ;{\rm Type-I}),\\
M_n &=& 2\pi n\ \ \ \ (\kappa^2 = 1/2\ ;{\rm BPS}),\\
M_n &>& 2\pi n\ \ \ \ (\kappa^2 > 1/2\ ;{\rm Type-II}).
\end{eqnarray}
These relations are analytically \cite{de vega, bogo} 
and numerically \cite{paul} investigated. 
Fig.\ref{fig:masssmall} and \ref{fig:vortexmass} showed 
the numerically solved $\kappa^2$-dependence of $M_n/n$, 
the vortex mass divided by the topological number $n$,
and we can see
\begin{eqnarray}
M_n < nM_1 \ \ \  (\kappa^2 <1/2), \nonumber \\
M_n > nM_1 \ \ \  (\kappa^2 >1/2).
\end{eqnarray}
These relations indicate
that the interaction between vortices is attractive for $\kappa^2<$1/2,
and repulsive for $\kappa^2>$1/2.
These are confirmed by the studies of the vortex-vortex interaction
analytically in large separation case \cite{kramer} and
numerically in arbitrary distance case \cite{jacob}. 

The qualitative understanding for the potential property 
is as follows.
The attractive part of the potential comes from
the reduction of the topological defect 
of the Higgs field condensate.
This is because
the total energy of the system decreases when
the domain of the normal conducting state decreases. 
On the other hand, the repulsive part comes from
the magnetic field interaction of the vortices.
Then, in the case of $\kappa^2<$1/2 where 
the coherence length of the Higgs field
exceeds the penetration depth of the magnetic field,
the attractive part becomes larger than the repulsive part
in the whole region,
and the vortex-vortex potential becomes attractive.
On the other hand, in the case of $\kappa^2>$1/2, 
the relation is reversed and the vortex-vortex potential
becomes repulsive.

In the case of the vortex-anti-vortex case,
the potential becomes attractive independently from $\kappa^2$
since the magnetic field interaction between the vortex
and the anti-vortex becomes attractive in contrast to 
the vortex-vortex case, and
there is no repulsive part in the potential.

Although
we consider the isolated vortex under the idealized situation
from the beginning,
it is worth mentioning under which situations
single vortex appears referring to the potential properties 
explained above.
In Type-I superconductor, the multi-vortices with total topological 
number $N$ transfer to the single, giant vortex with topological number $N$
due to their attractive interaction.
After the successive fusion processes,
the multi-vortices system transfers to the normal conducting state.
For the finite size superconducting materials
under the homogeneous magnetic fields and thermal equilibrium,
such fusion processes lead the difficulty to observe the
Type-I vortices because 
the magnetic fluxes successively invade from the interface between
the normal and superconducting phase.
Nevertheless, under the non-equilibrium situations or 
inhomogeneous magnetic fields concentrated on the small area,
it is expected that
the Type-I vortex or, more generally, the domain of the normal state
in the superconductor can appear as a transient state
in the process such as the dynamical evolution of the 
thermodynamic unstable state into the equilibrium state,
i.e., the nucleation or spinodal decomposition processes 
\cite{frahm, liu}. 
From the viewpoint of the particle physics,
an another interesting situation where Type-I vortex appears
is the case where
magnetic monopoles are immersed in superconductors
and the isolated vortex emerges between the monopole 
and anti-monopole \cite{Nambu}.
In this case, no collapse with the other vortices occurs
and Type-I vortex behaves as the stable object
due to the topological charge conservation.

On the other hand, in Type-II superconductor,
the giant vortex with topological charge $N$ fissions
to the $N$ vortices with charge
$n$=1 due to their repulsive nature \cite{jacob}.
This repulsive interaction prevents the vortices to 
coalescence and,
under the thermal equilibrium,
the $n$=1 vortices form some stable structure
like the Abrikosov triangle lattice \cite{abrikosov}.
Then, in Type-II superconductor, the $n$=1 case 
is realistic in the thermodynamic equilibrium
in contrast to Type-I case. 
However, when the system experience a certain quenching \cite{liu}
such as the rapid change of the temperature or density,  
the $n\ge 2$ vortex may be generated as the excited states.
This is the reason why we also consider the Type-II giant 
vortex in the following. 
Although they are unstable against the fission into small vortices,
when $\kappa^2$ is not too large,
it may be interesting to consider the vibrations of the giant vortex
because such vibrations can occur during the slow development
of the fission process due to
the small gradient of the vortex-vortex potential
for the small separation of vortices \cite{jacob}.
Moreover, the study of the fluctuation modes of the $n\ge 2$ vortex
may give some insight for the early stage of 
the evolution of the giant vortex into the multi-vortices.

%

\section{Axial-symmetric fluctuations}

In this section, using the results of the previous section,
we analyze axial-symmetric fluctuations
in the system which has only one vortex in the background. 
In subsection A, using the static vortex profile,
we derive the {\it vortex-induced potential} for 
the Higgs and photon field fluctuations,
and the equation of motion.
In subsection B, we closely discuss the potential
feature to classify the type of the excitation modes.

\subsection{Field equation for the fluctuations}
We investigate the axial-symmetric fluctuations 
of the Higgs field and the related photon fields around the 
static vortex.
We expand the fields with small amplitude fluctuations
around the static field,
\begin{eqnarray} \label{static}
\phi(t,z,r) = \phi^{ {\rm cl} }(r) + \varphi(t,z,r), \nonumber \\
A_\theta(t,z,r) = A^{ {\rm cl} }(r) + a_\theta(t,z,r).
\end{eqnarray}
Here we only consider the axial-symmetric fluctuations
and we include $t,z$-dependence at this stage.
It is important to notice that,
with the Ansatz (\ref{ansatz}) for the static solution, 
all the other type of fluctuations
such as fluctuations of $A_0,A_z,A_r$ and
the $\theta$-dependent fluctuations 
are {\it decoupled} from the axial-symmetric fluctuations which we consider.
Hence we can analyze the axial-symmetric fluctuations (\ref{static})
independently from the other type of fluctuations
and we drop off such fluctuation terms from the following Lagrangian.
The complete form of the Lagrangian is given in the Appendix \ref{appendixa}.

Substituting Eq.(\ref{static}) into Eq.(\ref{action2})
and using the static solution, 
the following expression for the 
axial-fluctuation part of Lagrangian can be derived as
\begin{eqnarray} 
\label{action4}
{\cal L}_{ {\rm axial} }^{\chi=0}[\phi, A_\theta] 
= {\cal L}_{ {\rm axial} }^{ {\rm cl} }
  + {\cal L}_{ {\rm axial} }^{(2)} 
  + {\cal L}_{ {\rm axial} }^{(3)} 
  + {\cal L}_{ {\rm axial} }^{(4)}.
\end{eqnarray}
The first term ${\cal L}_{ {\rm axial} }^{ {\rm cl} }$ represents 
${\cal L}_{ {\rm axial} }^{ {\rm cl} } 
\equiv {\cal L}_{ {\rm axial} }^{\chi = 0} [\phi_{ {\rm cl} },
A_{ {\rm cl} }]$.
The 1st order terms of fluctuations $\varphi$ and $a_\theta$
do not appear because we expand around the static solution.
The 2nd, 3rd, and 4th order fluctuation terms take the following forms;
\begin{eqnarray}
 {\cal L}_{ {\rm axial} }^{(2)} 
 \!\!\! &=& \!\!\! 
  (\partial_t \varphi)^2 - (\partial_z \varphi)^2 - (\partial_r \varphi)^2 
    \nonumber \\
 &&  
  + \frac{1}{2}(\partial_t a_\theta )^2
  - \frac{1}{2}(\partial_z a_\theta )^2
  - \frac{1}{2}(\partial_r a_\theta + \frac{ a_\theta }{r} )^2 
    \nonumber \\
 && 
  - V_{\varphi \varphi} \varphi^2 - V_{\varphi a} \varphi a_\theta 
  - V_{aa} a_\theta^2, \label{fluc} \\
{\cal L}_{ {\rm axial} }^{(3)} \!\!\! &=& \!\!\! 
    - 4 \kappa^2 \phi_{ {\rm cl} } \varphi^3 
    + 2 \bigg[ \frac{n}{r} - A_{ {\rm cl} } \bigg] \varphi^2 a_\theta
    -2 \phi_{ {\rm cl} } \varphi a_\theta^2, \\
{\cal L}_{ {\rm axial} }^{(4)} \!\!\! &=& \!\!\!
 - \kappa^2 \varphi^4 - \varphi^2 a_\theta^2. 
\end{eqnarray}
Here $V_{\varphi \varphi}$, $V_{aa}$, $V_{\varphi a}$
denote the {\it vortex-induced potential}, i.e.,
these interactions are induced by the static
vortex configuration which can be regarded as the 
large amplitude background external field.
$V_{\varphi \varphi}$ and $V_{aa}$ play roles as
the {\it single particle potential} 
for $\varphi$ and $a_\theta$ respectively,
and $V_{\varphi a}$ is the interaction which mixes the
$\varphi$ and $a_\theta$. 
Hereafter, we will simply call these interactions ``potentials''.
The concrete forms of the potentials are 
\begin{eqnarray}
&& V_{\varphi \varphi}(r)
   = \bigg[ \frac{n}{r}- A_{ {\rm cl} } \bigg]^2
   + 2 \kappa^2 ( 3 \phi_{ {\rm cl} }^2 - 1 ), \nonumber \\ 
&& V_{\varphi a}(r) = - 4 \bigg[ \frac{n}{r} - A_{ {\rm cl} } \bigg]
 \phi_{ {\rm cl} },\nonumber \\  
&& V_{aa}(r) = \phi_{ {\rm cl} }^2.
\end{eqnarray}
To investigate the fluctuation spectrum,
we solve the Euler-Lagrange equation for $ {\cal L}^{(2)} $
neglecting the higher order fluctuations. 
The quantum loop corrections can be calculated
using the Green's functions constructed of
bases which diagonalize $ {\cal L}^{(2)} $.
However, because of the inhomogeneous vortex-background,
the Green's functions are not functions of the distance
between arbitrary two points, i.e., $G(x,x')\neq G(x-x')$,
and the calculations of the loop corrections are not straightforward.
In addition, higher order terms of the axial-symmetric fluctuations 
couple with the terms of the non-axial symmetric ones.
We reserve the calculation of the loop corrections
for the future work
and, in this work, we safely neglect the quantum corrections
constraining ourselves to the case where the perturbative
expansion parameters $e^2$ and $e^2 \kappa^2=\lambda$
are sufficiently small.
(Recall that the perturbative corrections
are calculated with expansion of $e^{i\bar{S}}=e^{ie^2 S}$.
See also Eq.(\ref{reaction}).)
These constraints are not so strict because
the Higgs-photon coupling $e$ is usually small and we mainly
consider $\kappa^2\sim 1/2$ cases,
where $e^2 \kappa^2=\lambda$ is also sufficiently small.   

For later convenience to discuss the effect of the potential, 
we change the variables
\begin{eqnarray}
\left(
\begin{array}{c}
 \tilde{\varphi}(t,r,z) \\
 \tilde{a}_\theta(t,r,z) 
\end{array}
\right)
= r^{-1/2} \left(
\begin{array}{c}
\varphi \\
a_\theta / \sqrt{2} 
\end{array}
\right). \label{variable}
\end{eqnarray}
These changes of variables
enable us to considerably simplify the equation of motion as will be seen.
The Euler-Lagrange equation for 2nd order Lagrangian is obtained
through the variation,
\begin{eqnarray}
\frac{\delta}{\delta \tilde{\varphi}} \int dt dz dr
2\pi r {\cal L}^{(2)} 
= \frac{\delta}{\delta \tilde{a}_\theta} \int dt dz dr
2\pi r {\cal L}^{(2)} =0.
\end{eqnarray}
Then we obtain the following coupled equation
\begin{widetext}
\begin{eqnarray}
-\ (\ \partial_t^2 - \partial_z^2\ ) 
 \left[
\begin{array}{c}
 \tilde{\varphi} (t,z,r) \\
 \tilde{a}_\theta (t,z,r) 
\end{array}
\right]
&=& \left[
\begin{array}{cc}
- \partial_r^2 + \frac{ 4 ( n - A_{ {\rm cl} })^2 -1 }{4 r^2} 
	+ 2 \kappa^2 ( 3\phi_{ {\rm cl} }^2 -1 )   &
	-\frac{ 2 \sqrt{2} \phi_{ {\rm cl} } ( n - A_{ {\rm cl} })}{r} \\
  	-\frac{ 2 \sqrt{2} \phi_{ {\rm cl} } ( n - A_{ {\rm cl} }) }{r} &
	- \partial_r^2 + \frac{ 3 }{4 r^2} + 2 \phi_{ {\rm cl} }^2
\end{array} 
\right]
 \left[
\begin{array}{c}
 \tilde{\varphi} (t,z,r) \\
 \tilde{a}_\theta (t,z,r) 
\end{array}
\right] \\
&\equiv&
\left[
\begin{array}{cc}
- \partial_r^2 + V_{\alpha \alpha}(r) &
	V_{\alpha \beta}(r) \\
  	V_{\beta \alpha}(r) &
	- \partial_r^2 + V_{\beta \beta}(r)
\end{array} 
\right]
 \left[
\begin{array}{c}
 \tilde{\varphi} (t,z,r) \\
 \tilde{a}_\theta (t,z,r) 
\end{array}
 \right] .
\end{eqnarray}
\end{widetext}
%
\begin{figure*}[floatfix]
 \begin{minipage}{\linewidth}
     \includegraphics[width=15.0cm]{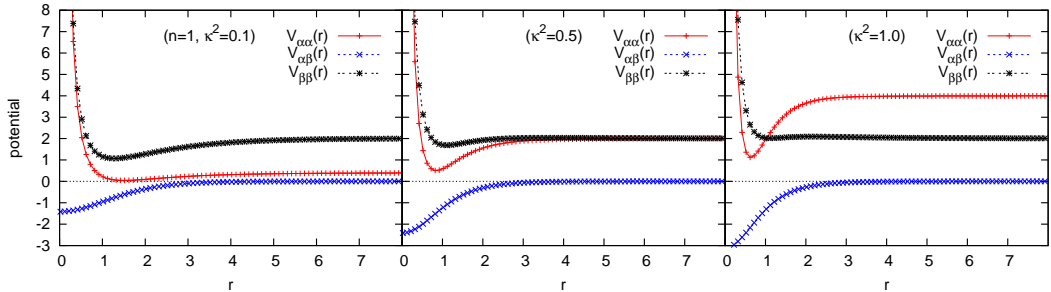}
 \vspace{-0.2cm}
 \caption{\label{potential1}
\small{ The plot of the {\it vortex-induced potential}
$V_{\alpha \alpha}(r), V_{\alpha \beta}(r), V_{\beta \beta}(r)$ 
as the {\it single particle potential} 
for fluctuations $\alpha(r)$ and $\beta(r)$ around single vortex,
  defined in Eq.(\ref{expansion}).
The Ginzburg-Landau parameter takes $\kappa^2$=0.1, 0.5, 1.0, and the
  topological number takes $n$=1. } }
\end{minipage}
\end{figure*}
Here $V_{\alpha \beta} (r) = V_{\beta \alpha} (r)$
(This is the consequence of the change of the variables 
in Eq.(\ref{variable})).
Note that the $t,z$ dependence of $\tilde{\varphi}$ and $\tilde{a}_\theta$
is trivial because the background static vortex configuration
is $t,z$-independent
and $t,z$ dependence does not appear in the potential. 
Then the partial differential equation can be 
decoupled into the total differential equations
for $t,z,r$, respectively.

The system has the translation invariance in both $z$-direction
and time direction, and then
the solution takes the following plane wave solution,
\begin{eqnarray}
\label{expansion}
\left(
\begin{array}{c}
 \tilde{\varphi}(t,r,z) \\
 \tilde{a}_\theta(t,r,z) 
\end{array}
\right)
= e^{-i(\omega t - k_z z) } \left(
\begin{array}{c}
 \alpha(r) \\
 \beta(r) 
\end{array}
\right).
\end{eqnarray}
Then the total differential equation in the radial direction for 
$\alpha(r)$ and $\beta(r)$ is obtained as
\begin{eqnarray}
m_j^2
\left[
\begin{array}{c}
 \alpha_j \\
 \beta_j 
\end{array}
\right] 
=
\left[
\begin{array}{cc}
- \frac{d^2}{dr^2} + V_{\alpha \alpha}(r) &
	V_{\alpha \beta}(r) \\
  	V_{\beta \alpha}(r) &
	- \frac{d^2}{dr^2} + V_{\beta \beta}(r)
\end{array} 
\right]
 \left[
\begin{array}{c}
 \alpha_j \\
 \beta_j 
\end{array}
\right], \nonumber \\
\label{psi1}
\end{eqnarray}
where $ m_j^2 \equiv \omega_{j,k_z}^2 - k_z^2 $.
Note that the Hermiticity of the matrix in the RHS 
in Eq.(\ref{psi1}) ensures the real value of $m_j^2$
and our later analysis shows $m_j^2>0$, i.e.,
the stability of the static profile against the fluctuations.
The root of the eigenvalue for the Eq.(\ref{psi1}), $m_j$,
plays a role of the mass for the 
propagating mode in the $z$-direction.
The Hermiticity of Eq.(\ref{psi1}) also enables us
to take the functions $\alpha(r)$ and $\beta(r)$ real,
and the orthonormal condition for the $\alpha(r)$ and $\beta(r)$ is
expressed as
\begin{eqnarray}
\int_0^{\infty}\!\!\!\! dr 
(\alpha_j,\beta_j) (\alpha_k,\beta_k)^T  
&=& \delta_{jk} \int_0^{\infty}\!\!\!\! dr\ [\alpha_j(r)^2 +
\beta_j(r)^2 ] \nonumber \\
&=& \delta_{jk}.
\label{psi2}
\end{eqnarray}

Here we give a brief explanation for 
the qualitative picture for the excitation modes.
The shape of the radial direction is given 
by $\alpha_j(r)$ and $\beta_j(r)$.
This wave function propagates in the $z$-direction with 
the frequency $\omega_{j,k_z}$ and the wave vector $k_z$.
Therefore
the root of the eigenvalue $m_j^2$ of the radial wave functions 
$\alpha_j(r)$ and $\beta_j(r)$
play roles of the mass for the propagation in the $z$-direction.
Then we call this type of new mode as the {\it peristaltic} mode 
of the vortex as illustrated in Fig.\ref{fig:peristic}.

\subsection{The feature of the potential }

We show in Fig.\ref{potential1} the single particle potentials
$V_{\alpha \alpha}$, $V_{\alpha \beta}$ and $V_{\beta \beta}$
around the vortex for various values of $\kappa$.
We can deduce the considerable information of the excitation modes
from the examination of these potentials.
To get the qualitative picture about the wave functions of fluctuations,
it is convenient to divide the region of the coordinate space $r$ 
into three regions, {\it central} ($0\sim0.5$), {\it intermediate}
($0.5\sim 4$), 
and {\it asymptotic} ($4\sim$) region in the rescaled Abelian Higgs model.

In the {\it central} region,
when $r \rightarrow 0$, 
the asymptotic behavior of the potential is
\begin{eqnarray}
&&V_{\alpha \alpha} \rightarrow \frac{4n^2 -1 }{4 r^2},\ \ \  
V_{\beta \beta} \rightarrow \frac{ 3 }{4 r^2} ,\ \nonumber \\
&&V_{\alpha \beta} = V_{\beta \alpha} \rightarrow 
({\rm negative\ const.})\times r^{n-1}\ . 
\end{eqnarray}
From these asymptotic behavior, 
it is expected that the fluctuations $\alpha(r)$ and $\beta(r)$ 
around the vortex center are
suppressed by the strong repulsive interaction 
proportional $1/r^2$ and
their distributions are shifted to the outside.

On the other hand, in the {\it intermediate} region,
the strong repulsive part of $V_{\alpha \alpha}$
and $V_{\beta \beta}$ becomes weak and the attractive
pocket emerges.
The mixing term $V_{\alpha \beta}$ also plays
an important role to enhance the mixing between 
the Higgs and photon field fluctuations, or $\alpha$ and $\beta$.
Since the $V_{\alpha \beta}$ always takes the negative value,
the excitation energy is reduced when
the relative sign of the $\alpha$ and $\beta$ fluctuations is positive.
To avoid the confusion,
we stress that the minus sign of $V_{\alpha \beta}$ 
does not mean attractive interaction.
Even if $V_{\alpha \beta}$ takes the positive value,
the energy is reduced when the relative sign of
$\alpha$ and $\beta$ fluctuations is negative.
This is also recognized from the fact that the replacement
 $(\alpha, \beta) \rightarrow (\alpha,-\beta)$ 
changes only the sign of the off-diagonal part of Eq.(\ref{psi1}).
In treatment up to the second order of the fluctuations,
only the absolute value of $V_{\alpha \beta}$ is relevant
to the mixing of $\alpha$ and $\beta$. 
All these features of the potentials 
$V_{\alpha \alpha},V_{\beta \beta},V_{\alpha \beta}$ make 
low energy modes of the fluctuations localized in the intermediate region. 

Finally we discuss the {\it asymptotic} region.
When $r \rightarrow \infty$, 
these potentials asymptotically behave as
\begin{eqnarray}
V_{\alpha \alpha} \rightarrow 4 \kappa^2,\ \ \  
V_{\beta \beta} \rightarrow 2,\ \ 
V_{\alpha \beta} = V_{\beta \alpha} \rightarrow	0. 
\end{eqnarray}
Note that the asymptotic behaviors do not depend on 
the topological charge $n$.
Since $V_{\alpha \beta}$ goes to zero,
the correlation between $\alpha$ and $\beta$
vanishes and these fluctuations behave as independent modes. 
Then we have only to consider the one-particle problem in this region, 
i.e., we treat the asymptotic forms of Eq.(\ref{psi1}) as
\begin{eqnarray}
(- \frac{d^2}{dr^2} + 4\kappa^2 ) \alpha_j(r) \simeq m_j^2 \alpha_j(r), \\
(- \frac{d^2}{dr^2} + 2 ) \beta_j(r) \simeq m_j^2 \beta_j(r).
\label{psi3}
\end{eqnarray}
The values $4\kappa^2$ and $2$ in the LHS play roles
of the (square of) {\it continuum thresholds} 
of the $\alpha$ and $\beta$ respectively.
These thresholds are equivalent
to the masses of the fluctuations 
in the system with no vortex, i.e.,
$2\kappa$ corresponds to the mass of the Higgs field
and $\sqrt{2}$ is the photon mass generated
through the Anderson-Higgs mechanism \cite{ander, higgs}.
Then, the threshold energy for the continuum states
is expressed as
\begin{eqnarray}
\label{thre}
m_{ {\rm th} }\equiv {\rm min}({2\kappa, \sqrt{2} } ).
\end{eqnarray}
When $m_j^2$ is smaller (larger) than the thresholds, 
the fluctuations behave like bound (continuum) states.

It is also worth mentioning that 
$V_{\alpha \alpha}$ and $V_{\beta \beta}$ take the same value
in the asymptotic region 
and the continuum thresholds are the same both for
$\alpha$ and $\beta$
for the case of $\kappa^2$=1/2,
where BPS saturation occurs.
This note is sometimes useful to roughly classify 
the type of the fluctuation spectrum in the low-energy region
in terms of the type of superconductor, as is explained below.
It is expected that, for Type-I case ($\kappa^2<$1/2), 
the threshold for the Higgs field is lower
than that for the photon field, and the Higgs field fluctuation 
dominates over the low energy physics.
For Type-II case ($\kappa^2>$1/2), the relation is reversed,
the photon field fluctuation dominates.
We also expect that the collective properties induced by 
the mixing of the Higgs and photon field fluctuations
are most pronounced 
when the BPS saturation occurs.
This is because the strength of both fluctuations
are the same order and they affect each other.
In the next section, we will see  
the above mechanism is crucial to emerge 
the characteristic discrete pole around $\kappa^2 = 1/2$. 

\begin{figure}[t]

     \includegraphics[width=8.5cm]{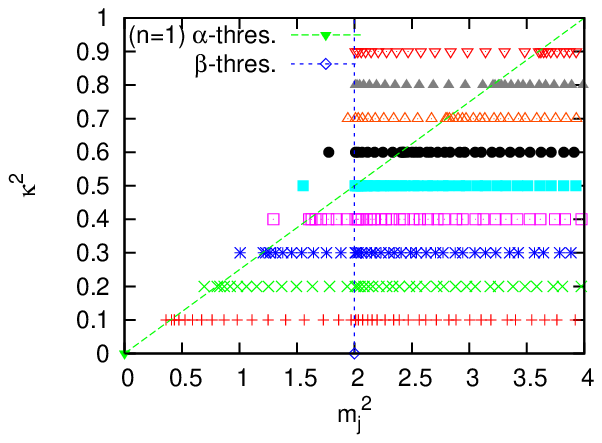}
    \vspace{-0.3cm}
    \caption{\small{ $\kappa^2$-dependence of $m_j^2$ for $n$=1 case 
 and the square of thresholds for $\alpha$ and $\beta$ fields,
 4$\kappa^2$ and 2, respectively.
 Around $\kappa^2$=$1/2$ (BPS value),
 as a collective mode of the Higgs and photon field fluctuations,
 there appear
 discrete poles in lower-energy region below (square of) 
 the continuum threshold
 $m^2_{ {\rm th} }={\rm min}(4\kappa^2, 2)$.
} }
 \label{fig:n1spec}
	  \vspace{0.2cm}
     \includegraphics[width=8.5cm]{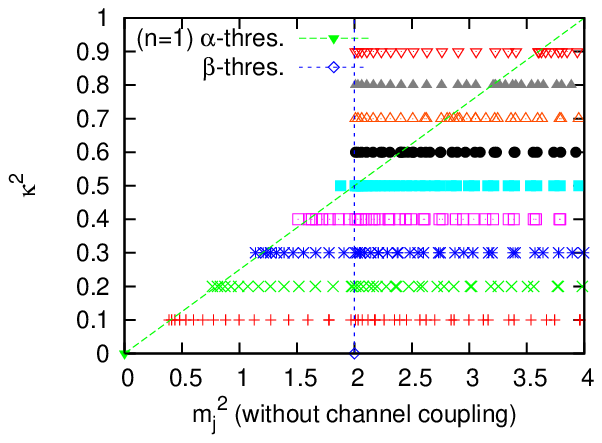}
    \vspace{-0.3cm}
    \caption{\small{ The same plot as Fig.\ref{fig:n1spec}
 except the replacement of $m_j^2$ with $m'^2_j$,
 i.e., $m_j^2$ calculated in the case that
 the channel coupling between Higgs and photon field fluctuations
 is artificially dropped off. 
 Without channel coupling, the bound-state with $\kappa^2 \simeq 1/2$
 becomes the shallow
 bound state due to the lack of the interplay between 
 the Higgs and photon field fluctuations. 
} }
 \label{fig:n1specw}
	  \vspace{0.2cm}
  \end{figure}
%
\section{Appearance of the discrete pole as the peristaltic mode of the
 vortex}
In this section, we summarize the numerical results
in both Type-I and Type-II superconductors.
The following results are obtained by solving the coupled channel 
equations (\ref{psi1}) numerically with the lattice spacing 0.02 and the 
lattice size 40.0 in our rescaled unit.
We impose the boundary condition on the
wave function $\alpha(r)$ and $\beta(r)$ to be zero value
at the box boundary.
In the following, we consider $k_z=0$ case.
The energy and radial wave functions shown in this section
are translated into those in the general $k_z \neq 0$ case
with the replacements $E=m_j \rightarrow (m_j^2 + k_z^2)^{1/2}$
and multiplying $e^{ik_z z}$ to the radial wave functions,
respectively.
\subsection{Analysis for the excitation spectrum}
In this subsection, we show the numerical results
for $n$=1 case in both Type-I and Type-II superconductors.
In Fig.\ref{fig:n1spec}, the $\kappa^2$-dependence of 
$m_j^2$ for $n$=1 case and the thresholds for $\alpha$ and $\beta$ fields are
plotted. 
Shown in Fig.\ref{fig:n1specw} are the same quantities as
Fig.\ref{fig:n1spec}, but without channel coupling,
which is artificially dropped off.

With smaller $\kappa^2 (<1/2)$, the threshold for $\alpha$ is shifted
lower than threshold for $\beta$, 
and $\alpha$ can excite with smaller energy and
the amplitude of $\alpha$ is enhanced compared to that of $\beta$ field.
Then the lower excitation modes are mainly characterized by
the $\alpha$ field fluctuation.
Under the normalization condition (\ref{psi2}),
there appears a large excess of 
the amplitude of $\alpha$ field compared to that of $\beta$ field,
and this mismatch between $\alpha$ and $\beta$
leads the suppression of the mixing effect.
Then the properties of the lower excitations asymptotically 
approach to those calculated without channel coupling.  
The property of the discrete pole
can be interpreted as the bound state
of the static vortex and the fluctuating 
Higgs field.

On the other hand, with larger $\kappa^2(>1/2)$,
the enhancement of the threshold energy for $\alpha$
leads the suppression of the $\alpha$ field excitation, 
and simultaneously enhances the $\beta$ field ratio
in the lower excitation modes.
Then the increasing of $\kappa^2$ value 
leads the reduction of 
the mixing correlation energy which is indispensable for the appearance of 
the discrete pole.
Eventually, at $\kappa^2 \simeq 0.8$, 
the discrete pole is buried 
in the continuum state spectrum of the fluctuating photon field
and no discrete pole appears for $\kappa^2 > 0.8$.
This indicates that such a pole is difficult
to be identified in most Type-II superconductors.

Finally we consider the most interesting case, $\kappa^2 \simeq 1/2$
case, i.e., near BPS saturation.
In this case, the mixing effect becomes larger,
and, as a result, the correlation between $\alpha$ and $\beta$
considerably lowers the excitation energy.
This is clearly seen from
the comparison of Fig.\ref{fig:n1spec} with
Fig.\ref{fig:n1specw}.
In Table \ref{tab:table1}, we summarize
the results for the lowest excitation energy $m_{j=0}$ and 
their relations to the threshold energy 
$m_{ {\rm th} } = {\rm min}\{2\kappa, \sqrt{2}\}$,
the lowest excitation energy without channel coupling, $m'_0$,
and the energy differences among them. 
The quantity $m_{ {\rm th} } - m_0$ represents the binding energy
and specifies
the spatial extension of the wave functions.
The quantity $m'_0-m_0$ is, roughly speaking,
the correlation energy induced by the mixing effect
between Higgs and photon field fluctuations.
As expected, the maximum mixing correlation
is realized in the BPS case ($\kappa^2$=1/2).

\begin{table}
\caption{\label{tab:table1} $\kappa^2$ dependence
of the various energy in the rescaled unit.
The physical value can be obtained by multiplying $v/e$.
$m_{ {\rm th} }$, $m_0$, $m'_0$ represent the
threshold energy, the lowest excitation energy
and the lowest excitation energy without
the channel coupling, respectively.
The negative sign of $m_0-m_{\rm th}$ indicates the 
bound state.
The quantity $m_0-m'_0$ roughly
characterizes the correlation energy of the 
Higgs-photon field mixing. }
\begin{ruledtabular}
\begin{tabular}{cccccccc}
\ $\kappa^2$ &\ \ \ \ \ \ $m_{{\rm th}}$ &\ \ \ \ \ $m_{0}$ 
&\ \ $m_0-m_{ {\rm th}}$
&\ \ \  $m'_{0}$ &\ \ \ $m_{0}-m'_{0}$  \\
\hline
\ 0.3 &\ \ \ \ \ \  1.095 &\ \ \ \ \ 1.003 &\ \ $-$0.092
&\ \ 1.068 &\ \ \ $-$0.065 \\
\ 0.4 &\ \ \ \ \ \  1.265 &\ \ \ \ \ 1.138 &\ \  $-$0.127
&\ \ 1.229 &\ \ \ $-$0.091 \\
\ 0.5 &\ \ \ \ \ \  1.414 &\ \ \ \ \ 1.247 &\ \  $-$0.167
&\ \ 1.371 &\ \ \  $-$0.124  \\
\ 0.6 &\ \ \ \ \ \  1.414 &\ \ \ \ \ 1.333 &\ \ $-$0.081
&\ \ 1.414 &\ \ \ $-$0.081 \\
\ 0.7 &\ \ \ \ \ \  1.414 &\ \ \ \ \ 1.394 &\ \ $-$0.020
&\ \ 1.414 &\ \ \ $-$0.020  & & \\
\end{tabular}
\end{ruledtabular}
\end{table}

\subsection{The spatial behavior of the fluctuations and
the character change of the excitation modes with $\kappa$}
\begin{figure*}[floatfix]
  \includegraphics[width=15.0cm]{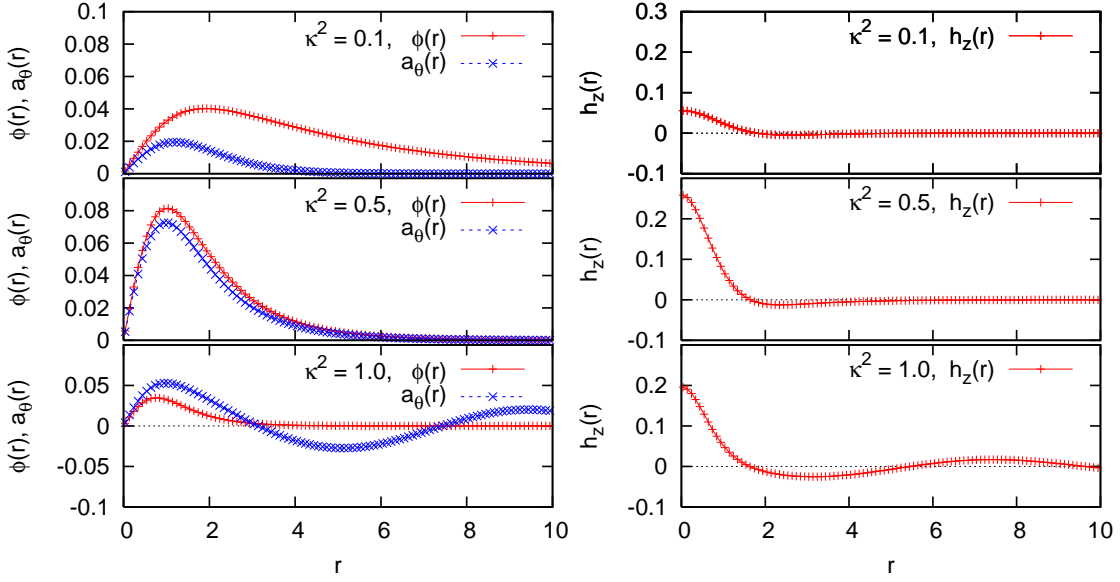}
 \vspace{-0.3cm}
 \caption{\small{ The behavior of the low-lying
mode of Higgs field fluctuation 
$\varphi(r) \equiv r^{1/2} \alpha(r)$,
the photon fluctuation $a_\theta(r) \equiv \sqrt{2} r^{1/2} \beta(r)$ and
the magnetic field fluctuation 
$ h_z(r) \equiv \frac{1}{r} \frac{d}{dr} (ra_{\theta}) $
for $n$=1 vortex.
For $\kappa^2 = 0.1,\ 0.5$ cases, we plot the lowest excitation mode
with the radial mass $m_0$ smaller than the threshold $m_{ {\rm th} }$.
For $\kappa^2 = 1.0$ case, where the discrete pole
does not appear, we plot the typical low-lying excitation mode
with the radial mass $m_j$, satisfying 
$ 2 < m_j^2 < 4 \kappa^2 $.} }
 \label{fig:n1wave}
	  \vspace{0.2cm}
\end{figure*}
\begin{figure}[t]
 \includegraphics[width=8.0cm]{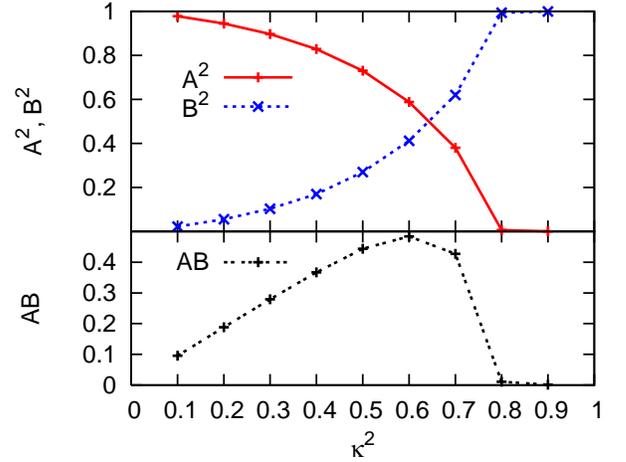}
 \vspace{-0.3cm}
 \caption{\small{ The $\kappa^2$-dependence of 
 $A^2 \equiv \int_0^{\infty} dr \ \alpha(r)^2$, 
 $B^2 \equiv \int_0^{\infty} dr \ \beta(r)^2$
and 
$A \cdot B \equiv \int_0^{\infty} dr \ \alpha(r) \beta(r)$
for the lowest 
excitation mode in the case of $n=1$.} }
\label{fig:n1mix}
\end{figure}
\begin{figure*}[floatfix]
     \includegraphics[width=15.0cm]{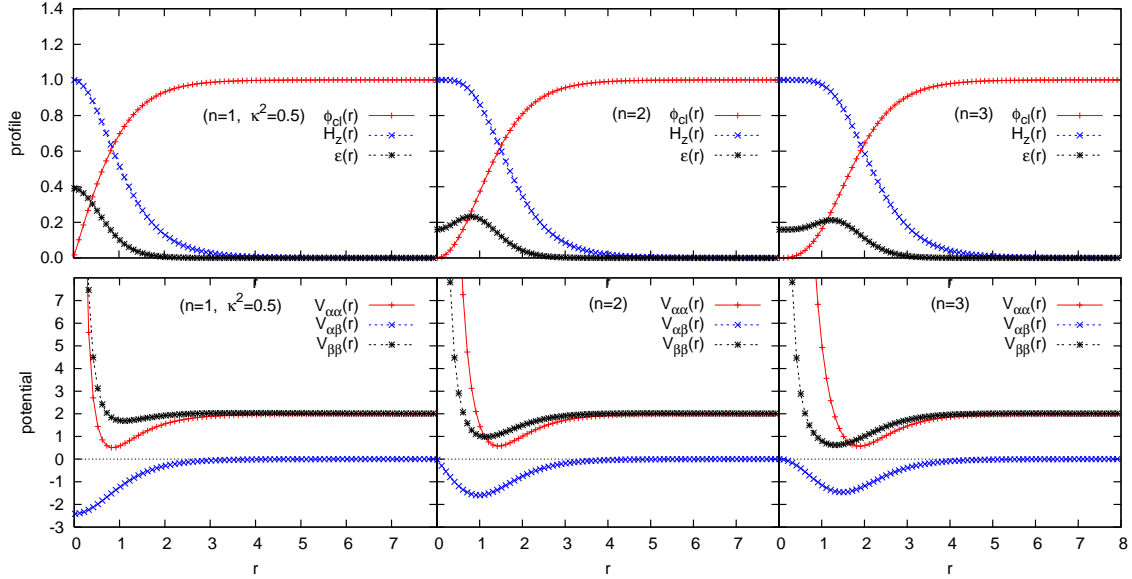}
 \vspace{-0.2cm}
 \caption{\label{potential2}\small{Shown in the upper panel
 are the profiles for $\phi_{ {\rm cl} }(r)$,
 the magnetic field $H_z(r)$ and static field energy ${\cal E}(r)$ 
 for $n$=1, 2, 3 vortices.
 The plots in the lower panel show the corresponding 
potential $V_{\alpha \alpha}(r), 
V_{\alpha \beta}(r)$\
 $V_{\beta \beta}(r)$
for the fluctuations $\alpha$ and $\beta$.
 $\kappa^2$ is fixed to the BPS value, i.e., $\kappa^2$=1/2. } }
\vspace{-0.2cm}
\end{figure*}
The features explained above can be explicitly seen from
the behavior of the radial 
wave function of the Higgs and the photon fields, i.e., 
\begin{eqnarray}
\varphi(r) = r^{1/2}\alpha(r),\ 
 a_\theta(r) = \sqrt{2}r^{1/2}\beta(r).
\end{eqnarray}
For the argument of the radial part, we hereafter drop off  
the trivial factor $e^{-i\omega t+k_zz}$ appearing in 
the relations (\ref{variable}) and (\ref{expansion}).
In fact, $\varphi(r)$ and $a_\theta(r)$
denote the radial part of the Higgs and the photon
fluctuations in Eq.(\ref{static}).

In Fig.\ref{fig:n1wave} we show the wave functions
$\varphi(r)$ and $a_\theta(r)$ of 
the low-lying modes and the corresponding magnetic fluctuations
\begin{eqnarray}
h_z(r) \equiv \frac{1}{r} \frac{d}{dr}(r a_\theta(r) ),
\end{eqnarray}
for $\kappa^2$= 0.1, 0.5, 1.0.
For $\kappa^2$ =0.1 and 0.5, 
we plot the lowest eigenfunction with radial mass smaller than
the threshold $m_{ {\rm th} }$.
On the other hand, 
for $\kappa^2$ = 1.0, there is no discrete pole
because the radial mass $m_j$ is larger than
the threshold $m_{ {\rm th} }=\sqrt{2}$,
i.e., $m_j^2>2$.
Then we plot the typical 
wave function of the low lying mode
whose radial mass satisfying
$2 < m_j^2 < 4\kappa^2$.

The behavior of the fluctuation fields 
coincides to our previous expectations.
For $\kappa^2$=0.1, 
the Higgs field fluctuation considerably exceeds to 
the photon fluctuation.
Although both of fluctuation fields have the smaller excitation energy
than the threshold $m_{{\rm th}}$ and
this leads localization of them around the vortex core,
the distribution of the Higgs field fluctuation spreads broader 
than that of the photon field fluctuation.
This is because the lowest excitation energy $m_0$ is close to 
the threshold for the Higgs field fluctuations, $2\kappa$,
and the lowest mode is characterized as
the shallow bound-state between the static vortex and the 
Higgs field fluctuation.

When we change to the $\kappa^2$=0.5, the mixing correlation
induces the corporative behavior of the Higgs and photon
field fluctuations and enlarges the binding energy $m_{ {\rm th} }-m_0$.
As a result, the lowest excitation mode becomes the deeper bound-state,
and then, the Higgs field localizes closer to the core, 
which affects the magnetic field vibration more strongly.

For $\kappa^2$=1.0,
because of $2 < m_j^2 < 4\kappa^2$ for low-lying modes,
only Higgs field fluctuation is localized and 
the photon field $a_\theta(r)$ oscillates asymptotically ($ r \rightarrow \infty$) as
\begin{eqnarray}
a_\theta(r) \sim e^{ik_r r}/r^{1/2},
\end{eqnarray}
leading to the asymptotic behavior of the magnetic field $h_z(r)$ like 
\begin{eqnarray}
h_z(r)=\frac{1}{r} \frac{d}{dr}(r a_\theta(r) )\sim e^{ik_r r}/r^{1/2}
\end{eqnarray}
with $k_r^2 \equiv m_j^2-2$.
(More precisely, $a_\theta(r)$ and $h_z(r)$
are expressed as the linear combinations of the
${\rm cos}(k_r r)/r^{1/2}$ and ${\rm sin}(k_r r)/r^{1/2}$ because
$\varphi(r)$ and $a_\theta(r)$ are real functions.)

Here we comment on the total magnetic flux.
In the static case, the total magnetic flux
takes the quantized value $2\pi n$, as seen in Eq.(\ref{totalflux}).
On the other hand, the photon fluctuation generally induces 
a time-dependent magnetic flux,
whose proper time-average vanishes due to the factor $e^{-i \omega t}$.
The contribution from the radial part of the magnetic field fluctuation is
\begin{eqnarray}
\delta \Phi \equiv 
\int_0^{\infty}\!\!\! dr\ 2\pi r h_z(r)
 &=& \int_0^{\infty}\!\!\! dr\ 2\pi \frac{d}{dr}( r a_\theta(r) ) \nonumber \\
 &=& 2 \pi \big[ r a_\theta (r) \big]^{r=\infty}_{r=0}.
 \label{totalflux2}
\end{eqnarray}
For $r\rightarrow 0$, $a_\theta(r)$ asymptotically approaches to zero.
On the other hand, for $r\rightarrow \infty$,
\begin{eqnarray}
r a_\theta(r) \rightarrow \left \{
\begin{array}{l}
C r^{1/2} e^{-\sqrt{m_{ {\rm th} }^2-m_j^2} r} \hspace{0.2cm}
({\rm discrete\ pole}),  \\
D r^{1/2} e^{ik_r r} \hspace{1.3cm} ({\rm continuum\ states}), 
\end{array}
\right.
\end{eqnarray}
where $C$, $D$ represent some constants.
Then, for the discrete pole, $\delta \Phi$ becomes zero
and thus
the total magnetic flux is conserved
in the arbitrarily cross section
perpendicular to the vortex line.
In contrast, for the continuum states,
the total magnetic flux becomes sensitive to
the situation far from the vortex core because of
the long tail behavior of $h_z(r)$.
For the continuum state, 
the radial photon fluctuation $a_\theta(r)$
asymptotically behaves as a two-dimensional spherical wave 
in the homogeneous no-vortex vacuum, and hence 
its asymptotic part can be regarded as a non-vortex-origin fluctuation. 
In other words, the continuum states can be regarded as the vortex plus a ``scattering wave''. 

To discuss the character change of the excitation modes
in wider range of $\kappa^2$, 
we define the functions
\begin{eqnarray}
A^2 \equiv \int_0^{\infty} dr \ \alpha(r)^2,\ \ \  
B^2 \equiv \int_0^{\infty} dr \ \beta(r)^2.
\end{eqnarray} 
Notice that $\int_0^{\infty} dr\ \big[ \alpha(r)^2 + \beta(r)^2
\big]=1$ in the  normalization condition. (See Eq.(\ref{psi2}).)
We also define the function to estimate the $\kappa^2$-dependence of the 
collective properties and their mixing ratio for the lowest 
eigenmode. 
One of the good indicators to estimate the degree of the
collective nature is the function 
\begin{eqnarray}
A \cdot B \equiv \int_0^{\infty} dr \ \alpha(r)
\beta(r). 
\end{eqnarray}
When the oscillation is sympathetic,
this integral takes the large value.
In Fig.\ref{fig:n1mix} we plot the behavior of the functions 
$A^2$, $B^2$ and $A \cdot B$ for various $\kappa^2$.
As expected, the ration of $\alpha(r)$, or Higgs field
decreases compared to that of $\beta(r)$, or
photon fluctuation with increasing $\kappa^2$
because the threshold for the Higgs field fluctuation increases.
Eventually, around $\kappa^2 \simeq 0.8$, the discrete pole
buries in the continuum, and the lowest mode
becomes the continuum state of the photon field. 
At the same time, the correlation between the Higgs and 
photon field fluctuations becomes very small,
as seen in the lower panel in Fig.\ref{fig:n1mix}.

\subsection{The peristaltic modes for $n \ge 2$ cases }
In this subsection, we show the results for the peristaltic modes
of the $n\ge 2$ vortex,
which is realistic in the case of $\kappa^2 \le 1/2$.
We emphasize the difference of the static vortex
between $n$=1 vortex and $n$=2, 3 vortex,
and then discuss their effects on the potential behavior
and the resulting excitation spectrum.
In this subsection, when we argue $n$-dependence of 
the classical profile and the potential,
$\kappa^2$ value is fixed to BPS value, i.e., $\kappa^2$=1/2.

Shown in Fig.\ref{potential2}
are the $n$=1, 2, 3 static vortex profiles and the corresponding
vortex-induced potentials
$V_{\alpha \alpha}$, $V_{\beta \beta}$, $V_{\alpha \beta}$
for the fluctuations $\alpha(r)$ and $\beta(r)$, respectively.
Here we qualitatively explain the changes of the static vortex profiles
in terms of $n$.
Since the total penetrating flux of the magnetic field is proportional 
to the topological number $n$, 
the magnetic field around the vortex core increases 
with increasing $n$.
Then the enlarged magnetic flux pushes the Higgs field
outward or reduce the expectation value of the Higgs field.
Roughly, this process spreads the normal phase region
near the vortex core and leads the outward shift of 
the interface between normal and superconducting phase.
According to the outward shift of the surface region,
where the kinetic energy of the Higgs field is enhanced,
the maximum of the local energy density also
shifts outward, as seen in the upper panel
of Fig.\ref{potential2}.

Next we present the qualitative explanation for 
the single particle potential for the fluctuation fields. 
First we give the explanation for the change of $V_{\alpha \alpha}$
for the Higgs field fluctuation.
Since the area of the magnetic flux is spread with increasing $n$,
the intersection surface of the static 
magnetic and the Higgs fields moves outward.
This leads the outward shift of the potential $V_{\alpha \alpha}$. 

The changes of $V_{\beta \beta}$ can be interpreted 
in somewhat indirect way as follows.
The fact read off from Fig.~\ref{potential2}
is the depth enhancement of the attractive pocket in $V_{\beta \beta}$
with increasing $n$.
This enables the photon field fluctuations to 
enhance and localize around the surface region 
more easily for larger $n$ cases.
The localization of the photon field fluctuation
means somewhat rapid spatial change of the wave function $a_\theta(r)$,
and leads a large fluctuation of the magnetic field
$h_z(r) = \frac{1}{r}\frac{d}{dr} (ra_\theta)$.
Then we can say that
the depth enhancement of the attractive 
pocket in $V_{\beta \beta}$ for large $n$
leads the large fluctuation of 
the magnetic field.
The large magnetic fluctuation reflects the reduction
of the Higgs field condensate
or the weakened Meissner effect.
In fact, the depth enhancement of the attractive pocket 
is closely related to the outward shift of the static Higgs 
field distribution.

Finally we comment on the potential $V_{\alpha \beta}$,
which induces the mixing correlation between
the Higgs and the photon field fluctuations.
As seen in Fig.~\ref{potential2},
with increasing $n$,
the maximum of $|V_{\alpha \beta}|$ shifts outward.
The physical meaning of the change in this potential
is naturally explained as the
outward shift of the interface between 
the static Higgs field and magnetic field
because, at the interface,
the fluctuations of the Higgs and photon field
can frequently interplay.

\begin{figure}[t]
\begin{minipage}{\linewidth}
     \includegraphics[width=8.5cm]{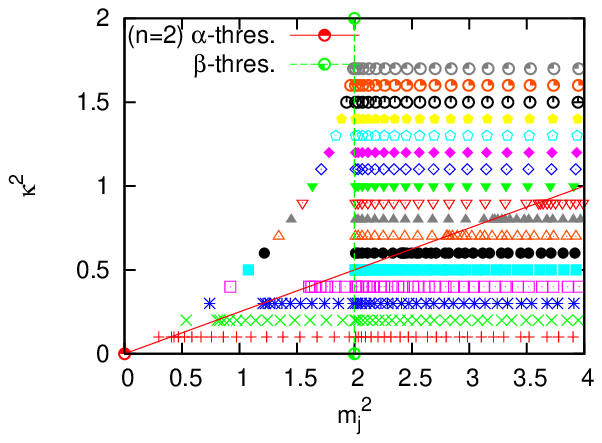}
    \vspace{-0.2cm}
    \caption{\small{ $\kappa^2$-dependence of $m_j^2$ for $n$=2 case 
 and the square of thresholds for $\alpha$ and $\beta$ fields, 
 $4\kappa^2$ and $2$, respectively.
 The excitation energy is considerably lowered than
 that in $n$=1 case and the discrete pole appears up to 
 considerably large $\kappa^2$.
} }
 \label{fig:nspe2}
	  \vspace{0.2cm}
\end{minipage}
\begin{minipage}{\linewidth}
 \includegraphics[width=8.5cm]{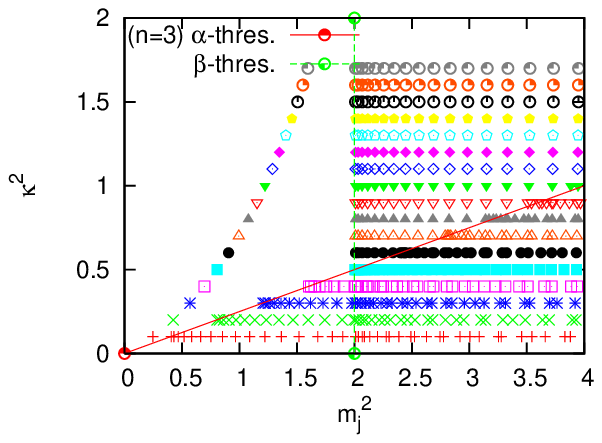}
 \vspace{-0.1cm}
 \caption{\small{ The same plot as Fig.\ref{fig:nspe2}
 except the replacement of the topological number $n$ from $n$=2
 to $n$=3.
 The thresholds for $\alpha$ and $\beta$ are unchanged.} }
 \label{fig:nspe3}
 \vspace{0.2cm}
\end{minipage}
\end{figure}

With increasing $n$, the resulting changes in the static configurations
and the vortex-induced potential lead the change of the excitation
energy of the peristaltic modes.  Now we see the excitation spectra of
the giant vortex ($n=2,3$) and their dependence on the $\kappa^2$ value.
The numerical results for $n=2,3$ cases are shown in
Figs.~\ref{fig:nspe2}, \ref{fig:nspe3}, respectively.  
For larger $n$,
the excitation energy for the lowest mode becomes much smaller as
expected from the arguments in the previous paragraphs. 
As a consequence of the depth enhancement of the attractive pocket in
$V_{\beta \beta}$, the discrete pole appears up to considerably large
$\kappa^2$ value in the Type-II region. 

In such large $\kappa^2 (>1/2)$ cases, of course, 
the fission processes of the giant
vortex to many $n$=1 vortices become important.
According to the analysis for the instability of the static vortex
\cite{bogo, gustafson}
and the interaction between two vortices \cite{jacob},
there exists no potential barrier for the fission from
single Type-II giant vortex to many $n$=1 vortices and
the giant vortex can fission through the non-axial
symmetric modes.
However, in spite of this observation,
we have a possibility that axial-symmetric discrete poles
play important roles
if the lifetime (decay width) of the giant vortex
is very large
compared to the typical time scale (excitation energy) 
of the axial-symmetric fluctuations.
The decay rate can be calculated through the analysis
for the non-axial-symmetric modes and the fluctuations of $A_0, A_r$,
which are not considered in this work.
We expect that, for $\kappa^2$ 
not so far beyond the BPS value, $\kappa^2=1/2$,
the lifetime is large because the force
to break up the giant vortex is small.
Then, if the lifetime is confirmed to be large enough,
we can immediately apply our results
without further consideration about 
the correlation between the axial-symmetric and
the other modes because these modes behave independently
as far as the distortion is sufficiently small, 
as noticed in Sec.IIIA. 
 
On the other hand, for $\kappa^2\le 1/2$ cases,
since the giant vortex is stable
and the axial-symmetric modes are independent from
the other modes not considered in this work, 
the strong bound-state-like modes 
more easily appear than the case for $\kappa^2>1/2$
because of non-existence of the fission processes.
In addition, according to the negative
surface tension, vortex with $\kappa^2\le 1/2$ tends to reduce the
surface area and then the axial-symmetric fluctuation becomes increasingly
prominent with smaller $\kappa^2$. 
Therefore, we expect that the discrete pole
shown in this section shares the considerable relevance to the
low-energy behavior of the giant vortex with $\kappa^2\le 1/2$. 
\section{Summary and Concluding Remarks}
In this paper, as a new-type mode beyond the stringy one,
we have studied the axial-symmetric
{\it peristaltic} modes of the vortex,
peculiar to the extension of the flux-tube.
We have investigated the peristaltic modes both in Type-I and Type-II
superconductors for $n=$1, 2, 3 and $\kappa^2=0.1\sim 1.0$.
We have treated single vortex in the situation where Type-I vortex can be 
realized without collapses with the other vortices
and behaves as the stable object.

Here we summarize our main results about the fluctuation modes.
With $\kappa^2 < 1/2$, the lowest excitation modes are
characterized by the Higgs field fluctuation binding with 
the static vortex.
On the other hand, with $\kappa^2 > 1/2$,
the mass of the Higgs field fluctuation becomes heavier and
the photon fluctuation modes are relevant.
For $\kappa^2 = 1/2$ (BPS cases), there appears the characteristic
discrete pole.
In this case, the Higgs and the photon 
field fluctuations have the same masses,
and then both fluctuations play important roles 
and their corporative behavior makes the lowest excitation softer one.
This mechanism is the main origin of the characteristic discrete pole.
Since the energy separation between this discrete pole 
and the continuum threshold is large around $\kappa^2 \simeq 1/2$,
the peristaltic mode may appear most clearly
in the superconductor with $\kappa^2 \simeq 1/2$.

We have also investigated the fluctuation modes of the
giant vortex with the topological number $n \ge 2$.
It is found that, with larger $n$, the photon field fluctuation
becomes more important for the lowest eigenmode
because the distribution of the static Higgs field is shifted outward
and the photon fluctuation can excite easily around the core.
We expect that, in $\kappa^2\le1/2$ cases,
the axial-symmetric excitation mode with the bound-state-like pole
plays an important role in the low-energy dynamics of the vortex.
On the other hand, for $\kappa^2>1/2$ cases,
although the equations for axial-symmetric excitation modes 
are decoupled from those of the other modes,
it is necessarily not only to investigate axial-symmetric modes
but also to consider whether these excitation modes can be realized
or not during the lifetime of the giant vortex.
To estimate the lifetime, 
we have to treat the non-axial-symmetric modes and 
the fluctuations of $A_0, A_r$, 
which drive the fission of the single giant vortex 
to many $n$=1 vortices.
We expect that, for $\kappa^2$ value not so far from the BPS value,
$\kappa^2$=1/2,
the life time is long enough and we can explore
the possibility of the appearance of the discrete pole. 

Finally, we end with a future perspective toward
the further investigation for the dynamics of the vortices
(or color-electric flux tubes) regarding them as dynamical degrees of freedom.

Although we restrict ourselves to the axial-symmetric static vortex 
profile and
fluctuations of $\phi$ and $A_\theta$,
we will investigate the more general cases without 
using the axial-symmetric Ansatz for 
the static profile and classify the large class of the fluctuation
modes, i.e., ($\phi$, $A_0$, $A_\theta$, $A_r$) in $\chi=0$ gauge,
or equivalently, ($\phi$, $\chi$, $A_\theta$, $A_r$) in $A_0=0$ gauge,
which are not fully considered in this work. 
For this purposes, we will calculate the static vortex profile and
excitation spectrum of the fluctuations  
in two-dimensional analysis in near future.

For Type-I case, we expect that, in low energy, 
the relevant excitation mode 
around single vortex is axial-symmetric since
the Type-I vortex tends to reduce the surface area
owing to the positive surface tension.
Also from the point of view on the kinetic energy,
the axial-symmetric excitation is favorable.
On the other hand, the arguments for the Type-II case are
more subtle.
Because Type-II vortex tends to increase the surface area
according to the negative surface tension,
it may be important to compare the surface energy to the 
kinetic energy of the distortion.

For $n\ge2$ giant vortices, interesting subjects are remained.
One of them is the calculation of the lifetime of single giant vortex
for $\kappa^2>1/2$ and the comparison to the excitation modes
considered in this work.
The lifetime of the giant vortex is closely related to
the dynamics of the multi-vortices in non-equilibrium systems.
As well as the information for the vortex-vortex potential,
the information for the lifetime of single giant vortex 
may provide the good building block to  
understand the interesting but complicated phenomena such as 
the transition of the system from 
the thermodynamically unstable state 
to the stable one.
In connection with them, we are also interested in 
the fluctuations between multi-vortices and
their evolution through the vortex-vortex scattering \cite{myers}.
As a successive study, 
we would like to investigate the results of the full two-dimensional 
analysis for excitation modes 
of the vortex in both single vortex and the multi-vortices system
\cite{kst}.

In application to QCD, the dynamics of the color-electric flux-tubes, 
or gluonic excitations as nonperturbative modes
are also important to discuss the hadrons, 
which are composed of quarks and color vortex between quarks.
In particular, the analysis for the multi-color-flux-tube system
can give an good insight to understand the color 
{\it confinement-deconfinement} transition in QCD, and properties of
the hot and dense matter of hadrons or quarks.
It is also interesting to consider heavy ion collisions
and the successive occuring expansion of the hot quark matter
in close analogy with the evolution of the early Universe.
The distribution of the topological defects may affect 
collective properties of quark gluon plasma and 
the resulting particle productions.

In this work, we have studied the dynamics of single vortex
as one example of the dynamics of the topological objects.
The treatments of topological objects as essential degrees 
of freedom are quite general and important to understand
the crucial aspects of the nonlinear field theory,
which sometimes can not be reached with the perturbative
treatments for the nonlinear terms.
We would like to extend our analysis to the other topological objects
such as color flux-tubes, instantons, 
magnetic monopoles in gauge theories,
or branes in the string theory \cite{nitta},   
and their many-body dynamics. 

\begin{acknowledgements}
The authors are grateful to Prof. R. Ikeda and Dr. K. Nawa 
for helpful discussions.
T.K. thanks to Drs. M. Nitta and E. Nakano for useful discussions
and suggestions in the YITP workshop, 
``Fundamental Problems and Applications
of Quantum Field Theory''.
H.S. was supported in part by a Grant for Scientific Research
(No.16540236, No.19540287) from the Ministry of Education,
Culture, Sports, Science and Technology (MEXT) of Japan.
This work is supported by the Grant-in-Aid for the 21st Century COE
``Center for Diversity and Universality in Physics" from the MEXT
of Japan.
\end{acknowledgements}

\vspace{1cm}

\noindent
{\bf Note added}: After this work is completed, we notice a similar 
study using a different gauge where ghosts appear \cite{GH95}.
Their numerical result on the physical modes is consistent with ours.

\appendix
\section{The decoupling of the axial-symmetric fluctuation modes}
\label{appendixa}
We give the complete form of the action
in the cylindrical coordinates ($r,\theta,z$).
We use the polar-decomposition of Eq.(\ref{polar}) 
for the Higgs field
\begin{eqnarray}
\psi(x) = \phi(t,r,\theta,z) e^{in\theta + i\chi(t,r,\theta,z)}
\ \  (-\pi\le\chi<\pi),
\end{eqnarray}
and fix the gauge to $\chi=0$. 
Then we obtain the action as
\begin{eqnarray} \label{action3}
 S^{\chi=0} = \int\!\! dt dr d\theta dz
\ r {\cal L}^{\chi=0}, 
\end{eqnarray}
where
\begin{widetext}
\begin{eqnarray} 
\label{lag3}
{\cal L}^{\chi=0} 
&=&  (\partial_t \phi)^2 - (\partial_r \phi)^2 
  - \frac{1}{r^2}(\partial_\theta \phi)^2- (\partial_z \phi)^2
 - \kappa^2 ( \phi^2 -1 )^2 - \frac{ (n-rA_\theta)^2 }{r^2} \phi^2  
  + (A_0^2-A_r^2-A_z^2)\phi^2 
  \nonumber \\
&&
  + \frac{1}{2} [E_r^2 + E_\theta^2 + E_z^2
                 - B_r^2 - B_\theta^2 - B_z^2].
\end{eqnarray}
Here the electric and magnetic fields are described as
\begin{eqnarray} 
\label{elemag}
   &&E_r = \partial_t A_r-\partial_r A_0, \ \ 
   E_\theta = \partial_t A_\theta - \frac{1}{r} \partial_\theta A_0, \
   \ 
   E_z = \partial_t A_z - \partial_z A_0, \\
 &&B_r = \frac{1}{r} \partial_\theta A_z - \partial_z A_\theta,\ \ 
  B_\theta = \partial_z A_r - \partial_r A_z, \ \ 
  B_z = \frac{1}{r} \partial_\theta (rA_r) 
  - \frac{1}{r}\partial_\theta A_r.
\end{eqnarray}
We decompose the field into the static part and fluctuation part,
\begin{eqnarray}
\label{flucap}
&& 
\phi(t,r,\theta,z) = \phi^{{\rm cl}}(r,\theta,z)
 + \varphi(t,r,\theta,z), \nonumber \\
&& 
A_{j}(t,r,\theta,z) = A^{ {\rm cl} }_{j}(r,\theta,z) +
 a_{j}(t,r,\theta,z),
\end{eqnarray}  
where $j=(0,r,\theta,z)$.
For the static part, we impose the Ansatz (\ref{ansatz}), or equivalently,
\begin{eqnarray}
\phi^{ {\rm cl} }(r,\theta,z) = \phi^{ {\rm cl} }(r), \ \ 
A^{ {\rm cl} }_\theta(r,\theta,z) = A^{ {\rm cl} }_\theta(r), \ \ 
A^{ {\rm cl} }_0 = A^{ {\rm cl} }_r = A^{ {\rm cl} }_z=0.
\end{eqnarray}
Putting these forms into the Lagrangian (\ref{lag3}),
we obtain
\begin{eqnarray}
{\cal L}^{\chi=0}[\phi, A_\theta] 
= {\cal L}^{ {\rm cl} }
  + {\cal L}^{(2)} + {\cal L}^{(3)} + {\cal L}^{(4)},
\end{eqnarray}
where
\begin{eqnarray}
&&{\cal L}^{ {\rm cl} }  
=  - (\partial_r \phi^{ {\rm cl} })^2 
   - \frac{1}{2}(\partial_r A^{ {\rm cl} }_\theta 
   + \frac{ A^{ {\rm cl} }_\theta }{r} )^2 
- \frac{ ( n- r A^{ {\rm cl} }_{\theta} )^2 } {r^2} 
    \phi_{ {\rm cl} }^2 
    - \kappa^2 (\phi_{ {\rm cl} }^2 -1)^2, \\
 &&{\cal L}^{(2)} 
=  (\partial_t \varphi)^2 - (\partial_r \varphi)^2 
  - \frac{1}{r^2}(\partial_\theta \varphi)^2- (\partial_z \varphi)^2
  - 2 \kappa^2 ( 3\phi_{ {\rm cl} }^2 -1 )^2 \varphi^2 
     - \frac{ (n-rA^{ {\rm cl} }_\theta)^2 }{r^2} \varphi^2  \nonumber \\
&& \ \ \ \ \ \ \ \ + \frac{1}{2} [e_r^2 + e_\theta^2 + e_z^2
                 - b_r^2 - b_\theta^2 - b_z^2]
+ \phi_{ {\rm cl} }^2 ( a_0^2 - a_r^2 - a_\theta^2 - a_z^2 )
  + \frac{ 4\phi_{\rm cl}( n - rA^{ {\rm cl} }_\theta )}{r} 
    \varphi a_{\theta}, \\
&&{\cal L}^{(3)} 
= - 4\kappa^2 \phi^{ {\rm cl} } \varphi^3
+ \frac{ 2(n-rA^{ {\rm cl} }_\theta ) }{r}  \varphi^2 a_\theta
+ 2 \phi^{ {\rm cl} } \varphi (a_0^2 - a_r^2 - a_\theta^2 - a_z^2), 
\\
&&{\cal L}^{(4)} 
=
\varphi^2 ( a_0^2 -a_r^2 -a_\theta^2 -a_z^2 ) - \kappa^2 \varphi^4. 
\end{eqnarray}
The Lagrangian for the static part, i.e., ${\cal L^{ {\rm cl} }}$ 
is related to
static energy (\ref{ene}) as ${\cal E} = -{\cal L}^{ {\rm cl} }$.  
The fluctuations of the electric and magnetic fields,
i.e., $(e_r,e_\theta,e_z)$ and $(b_r, b_\theta, b_z)$ are
expressed as the Eq.(\ref{elemag}) with the 
substitution $A_j \rightarrow a_j$.
The linearized Euler-Lagrange equations for the fluctuations
$\varphi, a_\theta, a_0, a_r, a_z$ are obtained as
\begin{eqnarray}
\!\!\!\!\!\!\!\!\!\!\!\!\!\!\!\!\!    
\bigg[
\frac{\partial^2}{\partial t^2}
-\frac{1}{r} \frac{\partial}{\partial r}
      \big(r \frac{\partial}{\partial r}\big)
-\frac{1}{r^2}\frac{\partial^2}{\partial \theta^2}
-\frac{\partial^2}{\partial z^2}
\bigg] \varphi 
 \!\!\!\!
&&= -\bigg[ \frac{(n-rA^{{\rm cl}}_\theta)^2}{r^2} 
    + 2\kappa^2(3\phi^{{\rm cl}}-1)^2 \bigg] \varphi 
+ \frac{2\phi^{{\rm cl}}(n-rA_\theta^{{\rm cl}})}{r} a_\theta,
\label{varphi} \\
\!\!\!\!\!\!\!\!\!\!\!\!\!\!\!\!\!    
\bigg[
\frac{\partial^2}{\partial t^2}
-\frac{1}{r} \frac{\partial}{\partial r}
      \big(r \frac{\partial}{\partial r}\big)
-\frac{\partial^2}{\partial z^2}
\bigg] a_\theta 
 \!\!\!\!\!\!
&&= -\bigg[\frac{1}{r^2}+ 2\phi_{{\rm cl}}^2 \bigg] a_\theta
 - \frac{4\phi^{{\rm cl}}(n-rA^{{\rm cl}}_\theta)^2}{r^2}\varphi
+ \frac{1}{r} \frac{\partial}{\partial \theta} \bigg[
   \frac{\partial a_0}{\partial t}
   -\frac{1}{r} \frac{\partial (r a_r)}{\partial r}
   -\frac{\partial a_z}{\partial z} \bigg], 
\label{atheta} \\
\!\!\!\!\!\!\!\!\!\!\!\!\!\!\!\!\!    
\bigg[
- \frac{1}{r} \frac{\partial}{\partial r}
      \big(r \frac{\partial}{\partial r}\big)
-\frac{1}{r^2}\frac{\partial^2}{\partial \theta^2}
-\frac{\partial^2}{\partial z^2}
\bigg] a_0 
\!\!\!\!\!\!
&&= -2\phi_{\rm cl}^2 a_0
 - \frac{\partial}{\partial t}
   \bigg[ \frac{1}{r^2} \frac{\partial a_{\theta}}{\partial \theta}
         + \frac{1}{r} \frac{\partial (r a_r)}{\partial r} 
	 + \frac{\partial a_z}{\partial z}\bigg], 
\label{a0} \\
\!\!\!\!\!\!\!\!\!\!\!\!\!\!\!\!    
\bigg[
 \frac{\partial^2}{\partial t^2}
 -\frac{1}{r^2} \frac{\partial^2}{\partial \theta^2}
 -\frac{\partial^2}{\partial z^2}
 \bigg] a_r 
\!\!\!\!\!\! 
&&= - 2\phi_{ {\rm cl} }^2 a_r
- \frac{2}{r^2} \frac{\partial a_\theta}{\partial \theta}
+ \frac{\partial }{\partial r}
   \bigg[ \frac{\partial a_0}{\partial t} 
   - \frac{1}{r}\frac{\partial a_\theta}{\partial \theta} 
   - \frac{\partial a_z}{\partial z} \bigg], 
\label{ar} \\
\!\!\!\!\!\!\!\!\!\!\!\!\!\!\!\!    
\bigg[
\frac{\partial^2}{\partial t^2}
-\frac{1}{r} \frac{\partial}{\partial r}
      \big(r \frac{\partial}{\partial r}\big)
-\frac{1}{r^2}\frac{\partial^2}{\partial \theta^2}
\bigg] a_z 
\!\!\!\!\!\! 
&&= -2\phi_{{\rm cl}}^2 a_z
 -\frac{\partial}{\partial z}
  \bigg[ \frac{\partial a_0}{\partial t}
          -\frac{1}{r} \frac{\partial a_\theta}{\partial \theta}
	  -\frac{1}{r} \frac{\partial (ra_r)}{\partial r} \bigg]
\label{az}.
\end{eqnarray}
\end{widetext}
It is important to notice that
we can decouple the equations for the 
axial-symmetric fluctuations of $\varphi$ and $a_\theta$
expanding the fluctuation modes
with the Legendre polynomial.
If we restrict ourselves to the $l=0$ modes analysis,
we can neglect angular-derivative terms
in Eq.(\ref{varphi})-(\ref{az}). 
Therefore, we can treat the fluctuations $\varphi$ and $a_\theta$
independently from fluctuations $a_0,a_r,a_z$,
because $\varphi$ and $a_\theta$ are related
to $a_0,a_r,a_z$ only through the angular-derivative terms.
As a result, to investigate the axial-symmetric fluctuations of 
$\varphi$ and $a_\theta$, 
we have only to treat the Lagrangian (\ref{fluc})
as mentioned in Sec.IIIA. 
\end{document}